\documentclass[a4paper]{aa}

\usepackage{natbib}
\bibpunct{(}{)}{;}{a}{}{,}
\usepackage{graphicx}
\usepackage{txfonts}
\usepackage{longtable}
\usepackage{lscape}
\usepackage{setspace}
\usepackage[labelfont=bf, nooneline]{caption}
\authorrunning{Letarte et al.}
\titlerunning{VLT/UVES Spectroscopy in Fornax GCs}

\begin{document}

\title{VLT/UVES Spectroscopy of Individual Stars in Three Globular
Clusters in the Fornax Dwarf Spheroidal Galaxy\thanks{Based on UVES
observations collected at the European Southern Observatory, proposal
number 70.B-0775}}

\author{ {\bf Bruno Letarte}\inst{1} \and {\bf Vanessa Hill}\inst{2}
\and {\bf Pascale Jablonka}\inst{3,4} \and {\bf Eline Tolstoy}\inst{1}
\and {\bf Patrick Fran\c cois}\inst{2} \and {\bf Georges
Meylan}\inst{4}}

\institute{
Kapteyn Astronomical Institute, University of Groningen, PO Box 800, 
9700AV Groningen, the Netherlands\\
\email{bruno@astro.rug.nl}, \email{etolstoy@astro.rug.nl} \and
Observatoire de Paris, GEPI CNRS UMR~8111, 2 pl.  Jules Janssen, 
92195 Meudon Cedex, France\\
\email{vanessa.hill@obspm.fr}, \email{patrick.francois@obspm.fr} \and
Observatoire de Gen\`eve, CH-1290, Sauverny, Switzerland \\
\email{pascale.jablonka@obs.unige.ch}\and
Laboratoire d'Astrophysique de l'Ecole
Polytechnique F\'ed\'erale de Lausanne (EPFL), 
CH-1290, Sauverny, Switzerland \\
\email{georges.meylan@epfl.ch}}

\date{Received October 28, 2005; accepted March 12, 2006}

\abstract{We present a high resolution (R$\sim$ 43~000) abundance
analysis of a total of nine stars in three of the five globular clusters
associated with the nearby Fornax dwarf spheroidal galaxy.  These three
clusters (1, 2 and 3) trace the oldest, most metal-poor stellar
populations in Fornax. We determine abundances of O, Mg, Ca, Ti, Cr, Mn,
Fe, Ni, Zn, Y, Ba, La, Nd and Eu in most of these stars, and for some
stars also Mn and La.  We demonstrate that classical indirect methods
(isochrone fitting and integrated spectra) of metallicity determination
lead to values of [Fe/H] which are 0.3 to 0.5 dex too high, and that
this is primarily due to the underlying reference calibration typically
used by these studies.  We show that Cluster~1, with [Fe /H]=$-2.5$, now
holds the record for the lowest metallicity globular cluster. We also
measure an over-abundance of Eu in Cluster~3 stars that has only been
previously detected in a subgroup of stars in M15. We find that the
Fornax globular cluster properties are a global match to what is found
in their Galactic counterparts; including deep mixing abundance patterns
in two stars.  We conclude that at the epoch of formation of globular
clusters both the Milky Way and the Fornax dwarf spheroidal galaxy
shared the same initial conditions, presumably pre-enriched by the same
processes, with identical nucleosynthesis patterns. 

}

\maketitle

\newcommand{\spr}{$s-$process}
\newcommand{\rpr}{$r-$process}
\newcommand{\alfe}{$\alpha$-elements}
\newcommand{\msun}{M$_\odot$}
\newcommand{\avgfeh}{$\langle$[Fe/H]$\rangle$}
\newcommand{\teff}{T$_{\mathrm {eff}}$}
\newcommand{\micro}{V$_{\mathrm {t}}$}
\newcommand{\vrad}{V$_{\mathrm {rad}}$}

\captionsetup[longtable]{font=small, justification=centering}
\captionsetup[table]{justification=centering}
\captionsetup[figure]{justification=justified}

\section{Introduction}

It is now established that some dwarf galaxies have globular
cluster systems
around them (Lotz et al.\citeyear{2004ApJ...613..262L}, van den Bergh
\citeyear{astro-ph/0509811}, Seth et al.
\citeyear{2004AJ....127..798S}). Their possible common origin with
clusters in  larger parent galaxies, the link between the dwarf galaxy
field and globular cluster stars are open questions to be addressed.
The largest samples of dwarf galaxies with globular cluster systems are
however distant, and this restricts the analyses to using integrated
properties.

\begin{figure}[!htb]
\begin{center}
\includegraphics[width=0.5\textwidth]{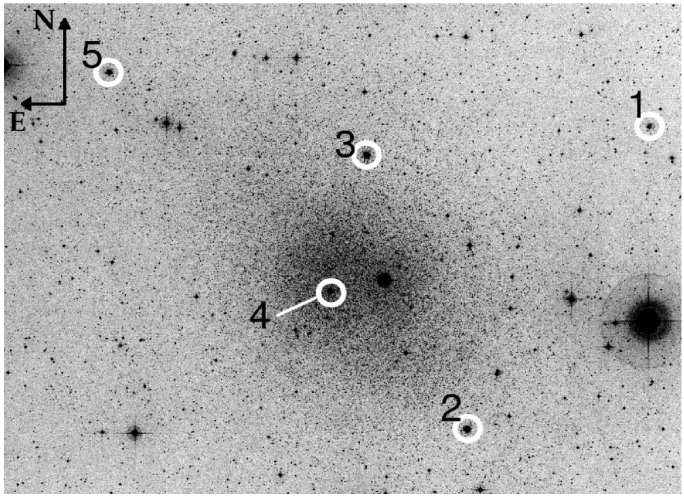}

\caption{A $\approx 85 \arcmin \times 62 \arcmin$ DSS image of the
Fornax dSph. North is up and East is to the left, as indicated. 
We have marked the
position of the 5 GCs using the numbering scheme defined by Shapley
\citeyear{1938NAT...142...715} and Hodge \citeyear{1961PASP...73Q.328H}.
\label{fig:dssfnx}}

\end{center}
\end{figure}

Fornax and Sagittarius are the nearest dwarf spheroidal galaxies (dSph)
with globular clusters and can be resolved into individual stars. The
Fornax dSph contains five star clusters (Shapley
\citeyear{1938NAT...142...715}; Hodge \citeyear{1961PASP...73Q.328H})
and while the Sagittarius dSph is obscured by dust and confused by
merging with our Galaxy, Fornax is high above the Galactic plane,
therefore offering a uniquely useful target for investigation, see
Figure~\ref{fig:dssfnx}.

The ages of the Fornax globular clusters have been determined by fitting
isochrones to deep HST Colour-Magnitude Diagrams [CMDs] (Buonanno et al.
\citeyear{1998ApJ...501L..33B}, \citeyear{1999AJ....118.1671B}). They
are found to be the same age as old metal poor Galactic clusters M92 and
M68 (around 13~Gyr old) to within $\pm$ 1~Gyr, with the exception of
Cluster~4, which seems buried in the center of Fornax and maybe younger
by about 3~Gyr.  The cluster metallicities have been estimated with
different techniques ranging from fitting a slope to the Red Giant
Branch (RGB) to high and medium resolution spectroscopy of the
integrated light of the cluster. Conclusions vary from one work to
another, as summarized in Strader et al.
(\citeyear{2003AJ....125.1291S}), but the clusters definitely appear
more metal-poor than the bulk of the galaxy field stellar population,
with bluer RGBs, well populated blue horizontal branches (HB) and a
range of HB morphology (Buonanno et al. \citeyear{1998ApJ...501L..33B}
and \citeyear{1999AJ....118.1671B}).  Saviane et al.
(\citeyear{2000A&A...355...56S}) showed that the Fornax dSph field star
colour distribution is well fitted by two Gaussian functions, best
interpreted as a bi-modal metallicity distribution, with the older
population having a wide abundance range between $-2.2$ and $-1.4$. 
Stars as young as $10^8$ Myr have also been discovered in the field
of Fornax (Stetson et al. \citeyear{1998PASP..110..533S}). In
this framework, the globular clusters of Fornax dSph trace the first
stages of star formation in the galaxy.

High resolution spectroscopy of individual stars in the clusters is the
only way to assess the abundances of individual chemical species.
Alpha, iron-peak, heavy -elements provide essential clues on (i) the
conditions of formation of the globular clusters in a dwarf galaxy,
including epoch and time scales (ii) to probe the nucleosynthesis in a
galactic system with a star formation history that is fundamentally
different from that of the Milky Way. We present here a VLT/UVES
spectroscopic analysis of a total sample of nine stars in three Fornax
dSph clusters.

\section{Observations}

We targeted Cluster~1, Cluster~2 and Cluster~3, to span the Fornax
globular cluster system range of distances from the galaxy centre
avoiding regions of heavy crowding. We also sample a range of HB
morphology, as well as the metallicity and concentration ranges.
Cluster~1, at a radial distance of 43 arcmin (or 1.75 kpc at the
distance of Fornax dSph) from the galaxy center, is diffuse, with low
surface brightness, most of its HB is red. Cluster~2, located at 25
arcmin (1~kpc) from the galaxy center, is slightly more concentrated and
exhibits a more extended HB.  Finally, Cluster~3 at a galactocentric
radial distance of 13 arcmin (530~pc) is very dense and has an extended
HB.

\begin{figure*}[!htb]
\begin{center}
\includegraphics[angle=0,width=1.0\textwidth]{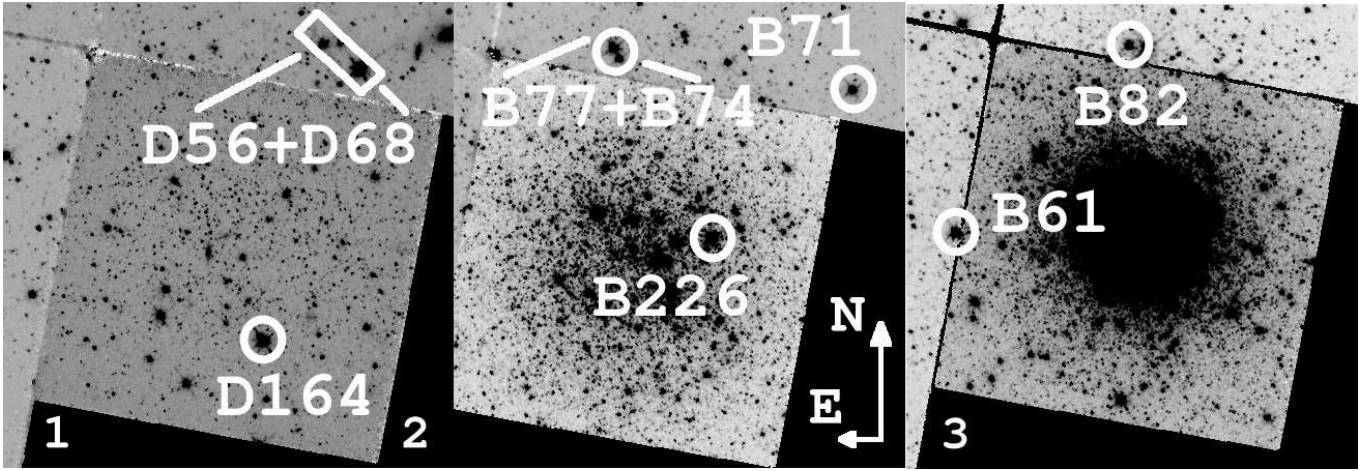}

\caption{The finding charts for our observations of
the Fornax GCs, from 1 (left) to 3
(right). North is up and East is left, as indicated.  
Note that star Cl3-B59
is outside of the cluster 3 HST field, to the west. \label{fig:fchart}}

\end{center}
\end{figure*}

We used the red arm of UT2/UVES, CD\#3, centered at 580nm, with a
wavelength range of 480-680nm (Dekker et al.
\citeyear{2000SPIE.4008..534D}) in visitor mode in October 2002.  We
obtained spectra with a resolution of $\sim$43~000 and average S/N
$\sim20-30$ per pixel with an integration time of $2-6$~hours for each
of the nine individual stars in Fornax dSph globular Clusters 1, 2 and
3.  The stars were selected to be on the RGB from CMDs, (Buonanno et al.
\citeyear{1985A&A...152...65B}; Demers et al.
\citeyear{1990PASP..102..632D}; Jorgensen \& Jimenez
\citeyear{1997A&A...317...54J} and Buonanno et al.
\citeyear{1998ApJ...501L..33B}).  Their individual finding charts are
shown in Figure~\ref{fig:fchart}.  We also observed 5 calibration red
giant branch stars in the well studied globular cluster M15 (Sneden et
al.  \citeyear{1997AJ....114.1964S}).  The observations of M15 stars
provide an independent check on our data reduction and analysis methods.
Details of the observations are shown in Table~\ref{tab:obslog},
including the derived radial velocities \vrad and S/N ratios.

\begin{table*}[!htb]
\begin{center}
\caption{Observation Log \label{tab:obslog}}
\begin{tabular}{lcrrrcrr}
\hline
\hline
Our id   &     Litterature IDs                      &   RA (J2000)&  DEC (J2000)&    exp. time&  S/N @ &  \vrad & Comments          \\
         &                                          &      degrees&      degrees&          (s)&  670 nm&  (km/s)&                   \\
\hline                                                   
Cl1-D56  &     D56$^b$, J24$^c$                     &    39.254958&    -33.17790&        14400&     23 &    57.6& same slit as D68  \\ 
Cl1-D68  &     B51$^a$, D68$^b$, J23$^c$            &    39.254609&    -34.17875&        14400&     50 &    60.2& same slit as D56  \\
Cl1-D164 &     B18$^a$, D164$^b$, J65$^c$, B713$^d$ &    39.257554&    -34.18628&        18000&     30 &    60.0&                   \\
Cl2-B71  &     B71$^a$                              &    39.677388&    -34.80412&        21600&     30 &    63.4&                   \\
Cl2-B74  &     B74$^a$                              &    39.684917&    -34.80301&        14400&     ...&    ... & Too faint         \\
Cl2-B77  &     B77$^a$                              &    39.685203&    -34.80303&        14400&     30 &    64.1& same slit as B74  \\
Cl2-B200 &     B200$^d$                             &          ...&          ...&         3900&     ...&    ... & carbon star       \\
Cl2-B226 &     B226$^d$                             &    39.682143&    -34.80801&         7200&     40 &    64.0&                   \\
Cl3-B59  &     B59$^a$, J9$^c$                      &    39.942803&    -34.25855&        10800&     30 &    59.7&                   \\
Cl3-B61  &     B61$^a$, J31$^c$                     &    39.957271&    -34.25782&        21600&     30 &    63.7&                   \\
Cl3-B82  &     B82$^a$, J3$^c$                      &    39.951057&    -34.25277&        14400&     40 &    64.8&                   \\
\hline
M15-S1   &     S1$^e$, K431$^f$                     &   322.484344&     12.21002&          900&    113 &  -106.4&                   \\
M15-S3   &     S3$^e$, K387$^f$                     &   322.481920&     12.21231&         1200&    115 &  -111.3&                   \\
M15-S4   &     S4$^e$, K825$^f$                     &   322.509599&     12.18986&          750&    122 &  -101.4& spec. double star \\
M15-S6   &     S6$^e$, K1040$^f$                    &   322.543661&     12.16832&         1000&     96 &  -100.0&                   \\
M15-S7   &     S7$^e$, K146$^f$                     &   322.457798&     12.13489&          900&     83 &  -100.7&                   \\
\hline
\multicolumn{2}{l|}{{ID$^a$ from \citet{1985A&A...152...65B}}} &  \multicolumn{3}{l|}{{ID$^c$ from \citet{1997A&A...317...54J}}} & \multicolumn{3}{l}{{ID$^e$ from \citet{1970ApJ...162..841S}}} \\
\multicolumn{2}{l|}{{ID$^b$ from \citet{1990PASP..102..632D}}} &  \multicolumn{3}{l|}{{ID$^d$ from \citet{1998ApJ...501L..33B}}} & \multicolumn{3}{l}{{ID$^f$ from \citet{1921VeBon..15....1K}}} \\
\end{tabular}
\end{center}
\end{table*}


\section{Data Reduction and Analysis}

The spectra were extracted with the standard UVES pipeline, except for
two pairs of stars on the same slit which we had to reduce interactively
using  the UVES context within MIDAS (see Table~\ref{tab:obslog}).  At
the telescope we already identified Cl2-B200 as a carbon star, and it
was discarded from further analysis.  As already noted by Sneden et al.
(\citeyear{1997AJ....114.1964S}),  M15-S4 is probably a spectroscopic
double star, as all lines are significantly wider (larger Full Width
Half Maximum [FWHM]) than the other stars of M15.  It was not used for
our abundance analysis.

For each of our targets we made equivalent width (EW) measurements with
SPLOT in IRAF, except for the lines with a small EW ($\lesssim$
50~m\AA).
For these weak lines, we noticed that SPLOT was giving very unstable
FWHM measurements.  A home-made gaussian-fitting program was used for
these lines to fix the FWHM at the instrumental value.  We also used
DAOSPEC\footnote {http://cadcwww.dao.nrc.ca/stetson/daospec/}, a new
programme that automatically measures EWs by iteratively fitting
gaussians of fixed FWHM to all lines in the spectrum and removing the
continuum signature (Stetson \& Pancino, in preparation).  Having
confirmed that DAOSPEC gives results compatible with those obtained by
hand for lines of moderate strength (EW $\leq$ 150 m\AA)\footnote{We
note here for completeness that, at this high resolution, the fixed FWHM
gaussian hypothesis adopted by DAOSPEC does not hold for the strongest
lines (EW $>$ 150 m\AA) where departures from the gaussianity and
natural broadening play a significant role.}, we used DAOSPEC measured
EWs for M15.

We can detect a range of elements in our coadded spectra: Fe~{\sc i},
Fe~{\sc ii}, Ti~{\sc i}, Ti~{\sc ii}, O~{\sc i}, Mg~{\sc i}, Ca~{\sc i},
Cr~{\sc i}, Mn~{\sc i}, Ni~{\sc i}, Zn~{\sc ii}, Y~{\sc ii}, Ba~{\sc
ii}, La~{\sc ii}, Nd~{\sc ii} and Eu~{\sc ii}, which allows us to
achieve a comprehensive abundance analysis. The most important is Fe,
with an average of 50 measured lines for Fe~{\sc i} and 10 lines for
Fe~{\sc ii}. Line parameters and EW measurements for all stars are
reported in Table~\ref{tab:LL_EW}.  Abundances for the different
elements were calculated with CALRAI, originally described in Spite
(\citeyear{1967AnAp...30..211S}) with many improvements over the years.
The stellar atmospheres models are those of Plez (private
communications, 2000 and 2002, described in Cayrel et
al.~\citeyear{2004A&A...416.1117C}).  Spectral synthesis was required
for some elements: Eu, Zn, Mg, Na, O and Ba to account for hyperfine
splitting (Eu, Ba); weak lines (Zn, O) and strong, possibility saturated
lines (Na, Mg).

\begin{figure}[!htb]
\begin{center}
\includegraphics[angle=270,width=0.5\textwidth]{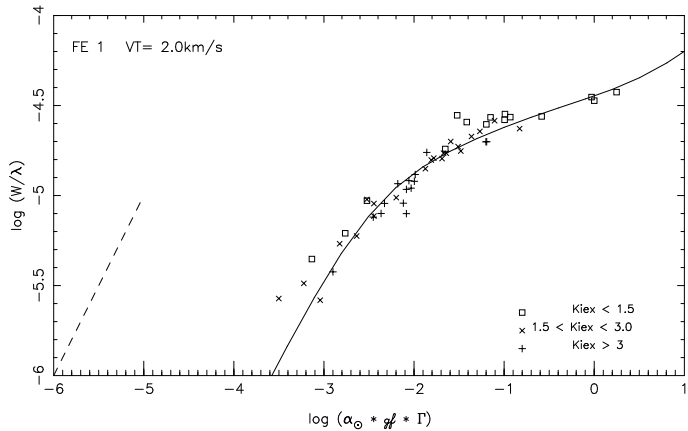}

\caption{Observed Curve of growth for Fe~{\sc i} in Cl3-B82. The dotted
line marks the [Fe/H]=0 location, while the full line is the theoretical
curve of growth for a typical Fe~{\sc i} line with the stellar
parameters adopted for this star.\label{fig:cog}}

\end{center}
\end{figure}

Initial guesses were made for the stellar effective temperature (\teff)
using $V-I$ and/or $B-V$ colours, using the Alonso et al.
(\citeyear{2001A&A...376.1039A}) calibration and a reddening of E($B-V$)
= 0.065. The surface gravity (log~g) was estimated assuming a $0.8
M_{\sun}$ mass for the stars, a distance modulus of (m-M)=20.85 mag and
bolometric corrections from Alonso et al.
(\citeyear{2001A&A...376.1039A}).  However, the quality of the
photometric data we gathered for these stars turned out to be too poor
to constrain firmly the star's effective temperature (only 2 stars had
HST photometry in Buonnano et al. \citeyear{1998ApJ...501L..33B}, and
the other photometric sources were ground-based, suffering from crowding
and not all in a homogeneous photometric system). We therefore chose to
base our analysis solely on spectroscopic criteria. The \teff, log~g and
micro-turbulence velocity (\micro) were adjusted to insure that we had
the ionisation balance of Fe~{\sc i} and Fe~{\sc ii} and that the
Fe~{\sc i} abundance is independent of both line strength and excitation
potential of the line.  Figure~\ref{fig:cog} illustrates the quality of
our solution by showing the curve of growth obtained for Cl3-B82, where
we notice that every part of the curve of growth is well populated.
The final set of stellar parameters used for each star are
shown in Table~\ref{tab:stelpar}.  The [Fe/H] in this table is the
metallicity of the model used to compute the abundances, not the final
abundance value of Fe~{\sc i} or Fe~{\sc ii} of the star.

As an additional test, since the S/N reached in the individual
spectra was rather limited, we also co-added the spectra of stars with
similar parameters within each cluster (all three stars of Cluster 3 on
the one hand, and the two cooler stars of Cluster~1 on the other hand),
and repeated the analysis. The results are fully consistent with the
analysis of the individual stars: \teff, log~g and \micro\ are
undistinguishable, while the mean [Fe/H] is recovered within 0.02 dex,
and most of the other abundance ratios fall well within the star to star
scatter.

\begin{table}[!htb] 
\begin{center} 
\caption{Adopted parameters of the stellar atmosphere model for each
star \label{tab:stelpar}}
\begin{tabular}{lrrrr} 
\hline 
\hline 
Star ID&    \teff (K)&   Log g&  [Fe/H]&  \micro (km/s)\\ 
\hline
Cl1-D56&         4600&     1.0&   -2.60&           2.1 \\ 
Cl1-D68&         4350&     0.5&   -2.60&           2.0 \\ 
Cl1-D164&        4400&     0.8&   -2.60&           2.1 \\ 
Cl2-B71&         4450&     0.7&   -2.10&           1.8 \\ 
Cl2-B77&         4350&     0.7&   -2.10&           1.7 \\ 
Cl2-B226&        4250&     0.6&   -2.10&           2.0 \\ 
Cl3-B59&         4400&     0.5&   -2.30&           2.0 \\ 
Cl3-B61&         4400&     0.8&   -2.30&           1.8 \\ 
Cl3-B82&         4350&     0.5&   -2.30&           2.0 \\ 
\hline                                            
M15-S1&          4350&     0.5&   -2.40&           1.9 \\ 
M15-S3&          4400&     0.6&   -2.40&           1.8 \\ 
M15-S4&          4150&     0.6&   -2.30&           2.3 \\ 
M15-S6&          4400&     0.7&   -2.40&           1.8 \\ 
M15-S7&          4400&     0.4&   -2.50&           1.9 \\ 
\hline 
\end{tabular}
\end{center} 
\end{table}

\begin{figure}[!htb]
\begin{center}
\includegraphics[angle=0,width=0.5\textwidth]{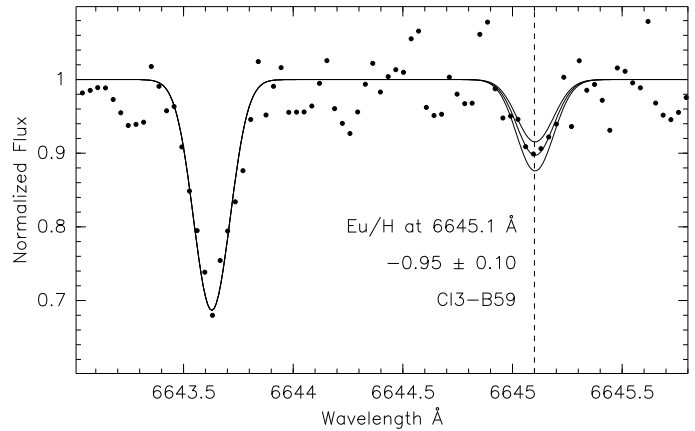}

\caption{The Synthetic spectra for the Eu line at $\lambda$ = 6645.1
$\AA$ overlaid on the data for Cl3-B59.  The middle line is the adopted
fit, while the lower and upper ones are the error estimate of $\pm$~0.1
dex. The larger line on the left is a Ni line.\label{fig:synthesis}}

\end{center}
\end{figure}

A significant source of error in our analysis is the uncertainty in
measuring the EW.  Our error in the EW determinations were estimated by
propagating the EW error estimates (from splot) through the abundance
computation (the abundances $\mathrm EW+\delta EW$ and $\mathrm
EW-\delta EW$ were computed and compared to the central adopted value).
For elements which were computed by spectral synthesis, the error is
estimated by eye, plotting a range of acceptable fits, as illustrated in
Figure~\ref{fig:synthesis} for the weak Eu line in Cl3-B59. Another way
to estimate the measurement errors affecting the abundance is to
consider the dispersion (rms) around the mean. For species with
sufficient number of lines measured ($>$3), this dispersion was adopted
whereas the direct measurement error was used for species probed by
fewer lines.  

However, there is more than just the measurement error to consider.  The
chosen stellar model will also affect the derived abundances.  The three
important parameters in the model are: temperature, gravity and
micro-turbulence velocity.  Each of these influences the final abundance
in a different way.  We estimated the uncertainty in each of these three
parameters using the corresponding statistical errors on the slopes
and the (Fe~{\sc i} - Fe~{\sc ii}) difference and computed the
resultant change in abundance for all elemental ratios.
Table~\ref{tab:errors} shows the abundance offset generated by each
parameter, and the combined effect of all three (added quadratically).
\begin{table}[!htb]
\begin{center}
\captionsetup{width=0.4\textwidth}
\caption{Dependencies on model atmosphere parameters \label{tab:errors}}
\begin{tabular}{lrrrr}
\hline
\hline
                             &   $\Delta$ \teff &  $\Delta$ Log~g & $\Delta$ \micro & Combined\\
                             &            -200 K&             -0.3& -0.2 km s$^{-1}$&         \\
\hline
$\textrm{[Ba~{\sc ii}/Fe~{\sc i}]}$                       &  -0.11  &  0.11  & -0.06  & 0.17\\
$\textrm{[Ca~{\sc i}/Fe~{\sc i}]}$                        &   0.00  & -0.01  &  0.05  & 0.05\\
$\textrm{[Cr~{\sc i}/Fe~{\sc i}]}$                        &   0.21  & -0.03  & -0.02  & 0.21\\
$\textrm{[Eu~{\sc ii}/Fe~{\sc i}]}$                       &  -0.16  &  0.13  &  0.08  & 0.22\\
$\textrm{[Fe~{\sc i}/H]}$                                 &   0.27  & -0.05  & -0.10  & 0.29\\
$\textrm{[Fe~{\sc ii}/H]}$                                &  -0.05  &  0.07  & -0.06  & 0.10\\
$\textrm{[La~{\sc ii}/Fe~{\sc i}]}$                       &  -0.12  &  0.12  &  0.08  & 0.19\\
$\textrm{[Mg~{\sc i}/Fe~{\sc i}]}$                        &  -0.06  & -0.08  &  0.02  & 0.10\\
$\textrm{[Mn~{\sc i}/Fe~{\sc i}]}$                        &  -0.02  & -0.01  &  0.08  & 0.08\\
$\textrm{[Na~{\sc i}/Fe~{\sc i}]}$                        &   0.18  & -0.04  & -0.04  & 0.19\\
$\textrm{[Nd~{\sc ii}/Fe~{\sc i}]}$                       &  -0.14  &  0.10  &  0.04  & 0.18\\
$\textrm{[Ni~{\sc i}/Fe~{\sc i}]}$                        &   0.02  & -0.02  & -0.03  & 0.04\\
$\textrm{[O~{\sc i}/Fe~{\sc i}]}$                         &  -0.11  &  0.13  &  0.08  & 0.19\\
$\textrm{[Ti~{\sc i}/Fe~{\sc i}]}$                        &   0.27  & -0.02  &  0.02  & 0.27\\
$\textrm{[Ti~{\sc ii}/Fe~{\sc i}]}$                       &  -0.21  &  0.09  &  0.02  & 0.23\\
$\textrm{[Y~{\sc ii}/Fe~{\sc i}]}$                        &  -0.18  &  0.09  &  0.04  & 0.21\\
$\textrm{[Zn~{\sc i}/Fe~{\sc i}]}$                        &  -0.28  &  0.03  &  0.06  & 0.29\\

\hline
\end{tabular}
\end{center}
\end{table}


A summary of our abundance analysis is available in
Table~\ref{tab:ratios_fnx} (Fornax) and Table~\ref{tab:ratios_m15} (M15)
where we present all of our elements with the associated error
estimates and the number of lines used to compute the ratio.
Only the EW measurement error is used in plots and tables, and
Fe~{\sc i} is used to determine our [el/Fe] ratios.

\section{Interpretation}

\subsection{The Iron abundance}

\begin{figure}[!htb]
\begin{center}
\includegraphics[width=0.5\textwidth]{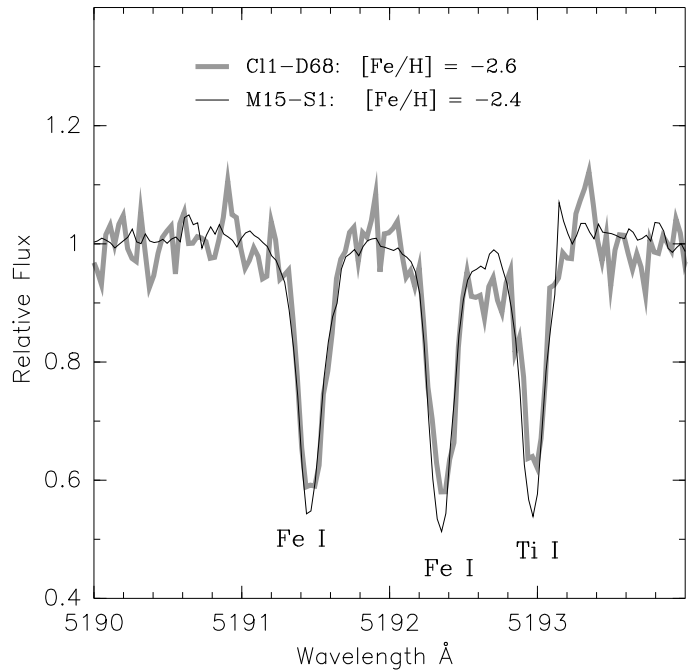}

\caption{The comparison between Cl1-D68 and M15-S1.
These are two RGB stars with similar
stellar parameters but a difference in [Fe/H] of 0.2 dex.
\label{fig:2spectra}}

\end{center}
\end{figure}

Table~\ref{tab:fnxfeh} compares our mean [Fe/H] with the latest results
of two different classical methods: RGB slope fitting and integrated
spectroscopy.  Our abundances appear 0.3 to 0.5 dex lower than previous
estimates.  Most of this discrepancy is attributable to different
reference calibrators.  Indeed, both the integrated spectroscopy and
isochrone fitting are based on the Zinn \& West
(\citeyear{1984ApJS...55...45Z}) metallicity scale which places M15 at
\avgfeh = $-2.15$ and M92 at \avgfeh = $-2.24$.  In contrast, high
resolution spectroscopic analyses consistently find \avgfeh =$-2.4$ for
M15, including  Sneden et al.  (\citeyear{1997AJ....114.1964S}) and this
present work.  Meanwhile, M92 is found to be \avgfeh =$-2.34$ (Sneden et
al., \citeyear{2000AJ....120.1351S}). The difference between high
resolution spectroscopy and the other indirect methods, due to
differences in calibration, is therefore of the order of 0.25 dex, the
rest of the discrepancy might be due to the propagation of errors, and
indeed appears of the order of the quoted error bars ($\pm$ 0.2dex). In
conclusion, although the absolute value of metallicities presented in
the works quoted in Table~\ref{tab:fnxfeh} do not appear accurate, the
comparison made by the authors with M15 and M92, the most metal-poor
clusters known in our Galaxy, is correct.  Our analysis reveals that
Cluster~1, at \avgfeh  = $-2.5$, is actually the most metal-poor
globular cluster yet observed.  It is clearly more metal-poor than M15,
with weaker iron lines, as can be seen in Figure~\ref{fig:2spectra},
where we compare Cl1-D68 and M15-S1, two RGB stars of similar 
temperature, surface gravity and micro-turbulence velocity.

\begin{table*}[!hbt] 
\begin{center} 
\caption{Recent metallicity estimates from different methods \label{tab:fnxfeh}}
\begin{tabular}{rrr|rr} 
\hline 
\hline 
Cluster 1& Cluster 2& Cluster 3& Method& Reference\\ 
\hline
$-2.5  \pm 0.1$&   $-2.1  \pm 0.1$&    $-2.4  \pm 0.1$&  Individual stars, HR spectra&   This work\\
             N/A&  $-1.76 \pm 0.41$&   $-1.84 \pm 0.18$&     Integrated light spectra&   Strader et al. \citeyear{2003AJ....125.1291S}\\
$-2.20 \pm 0.20$&  $-1.78 \pm 0.20$&   $-1.96 \pm 0.20$&                    RGB Slope&   Buonanno et al. \citeyear{1998ApJ...501L..33B}\\
\hline 
\end{tabular} 
\end{center} 
\end{table*}


\subsection{The Alpha elements}

Alpha elements come predominantly from Type II supernovae, unlike Fe
which comes predominantly from type Ia SN
(McWilliam~\citeyear{1997ARA&A..35..503M},
Tinsley~\citeyear{1979ApJ...229.1046T}).  The [$\alpha$/Fe] ratios
frequently display different patterns with respect to Fe in different
environments (e.g., Shetrone et al.  \citeyear{2001ApJ...548..592S}).
They are typically overabundant by +0.3 to +0.4 dex in Galactic globular
cluster stars and halo stars with respect to solar, as expected in old
components where only SNe~II have had time to contribute to the chemical
enrichment.

\begin{figure}[!htb]
\begin{center}
\includegraphics[angle=0,width=0.5\textwidth]{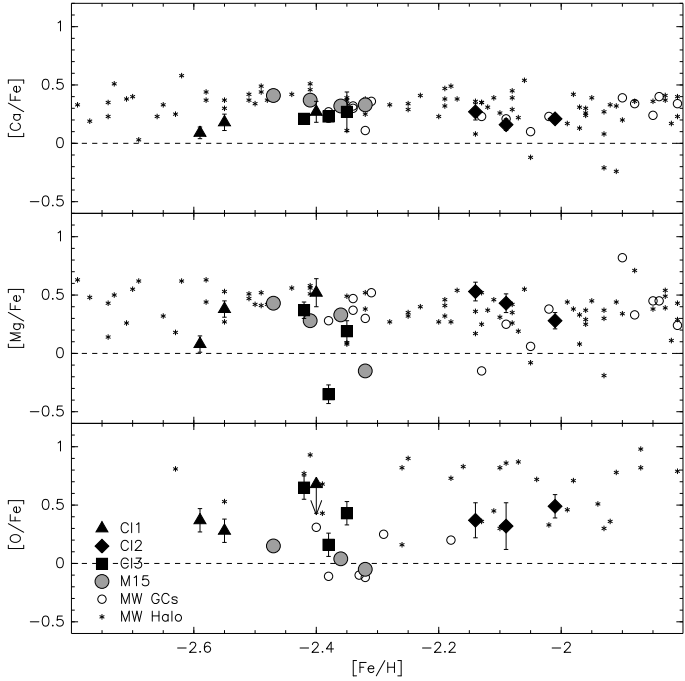}

\caption{Alpha elements abundances as a function of [Fe/H]. Filled
triangles are for Cluster 1, filled diamonds are for Cluster 2 and
filled circles are for Cluster 3.  Asterisk are for our M15 stars.
Small grey dots are galactic stars and small empty circles are galactic
GCs. Upper limits, when present, are shown with one sided arrows,
replacing the error bars.  See text for more details.
\label{fig:alpha_feh}}

\end{center}
\end{figure}

In Figure~\ref{fig:alpha_feh}, we plot the abundance ratios for \alfe\
Ca, Mg and O in Fornax dSph globular clusters 1, 2 \& 3.  Also plotted
are the four M15 control stars and, as smaller symbols, Galactic halo
stars, taken from the compilation of Venn et al.
(\citeyear{2004AJ....128.1177V}) and Galactic globular cluster stars
from the compilation (averaged by cluster) of Pritzl et al.
(\citeyear{2005AJ....130.2140P}), except for [O/Fe] points, which
are from Shetrone et al.~\citeyear{1996AJ....112.1517S} (individual
stars, not averages.) The abundances of Ca, Mg and O are all above the
solar value (ust like the Galactic halo and globular cluster stars) with
a small dispersion and small error bars.  There are a couple of Fornax
dSph globular cluster stars with clearly anomalous O and Mg abundance,
and they will be discussed later in section~\ref{subsec:deepmix}.  The
Fornax dSph globular cluster $\alpha$-element ratios appear to follow
the same patterns found in Galactic globular cluster stars, suggesting
that the oldest epoch of globular cluster formation is very similar in
these two different environments. The overabundance of \alfe\ seen in
Galactic globular clusters stars may be interpreted as the number of
massive stars present in the early history of our Galaxy assuming that
the main contributor to \alfe\ is SNe~II explosions from massive stars.
The same over abundance is seen in Fornax dSph globular clusters so this
enrichment pattern is not only present in our Galaxy.

\begin{figure}[!htb]
\begin{center}
\includegraphics[angle=0,width=0.5\textwidth]{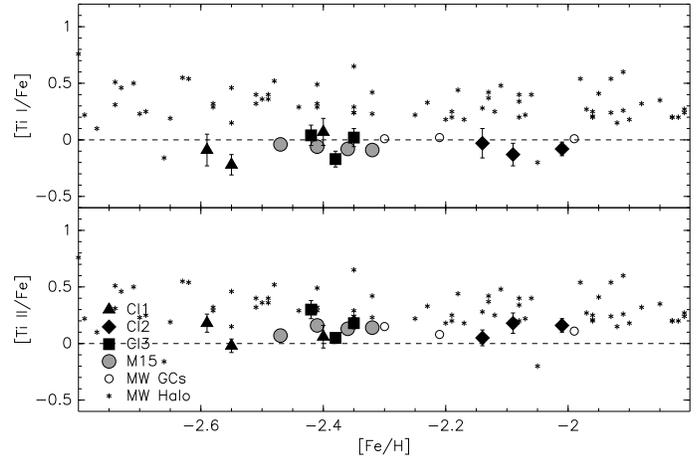}

\caption{Titanium abundances as a function of [Fe/H]. 
The symbols are the same as in Figure~\ref{fig:alpha_feh}. Seperated 
Ti~{\sc i} and Ti~{\sc ii} were not available for our halo stars, so a
global [Ti/Fe] is used for these points.
\label{fig:titan_feh}}

\end{center}
\end{figure}

Titanium is shown in Figure~\ref{fig:titan_feh}, where we chose
to compare our results with the Galactic globular clusters
studied by Shetrone et al. (\citeyear{2003AJ....125..684S}) rather than
the compilation of Pritzl et al. (\citeyear{2005AJ....130.2140P}) for
homogeneity purposes. At first glance, Ti seems to be underabundant in
the Fornax clusters with respect to halo stars in the Milky Way.
However, they fall right on top of our M15 and Shetrone's M30, M68 and
M55, close to a solar Ti/Fe ratio. However, we would like to stress that
[Ti/Fe] ratios of different authors can be on different scales
(depending on the set of Ti lines used and the adopted log~gfs), as well
illustrated by M15: the [Ti/Fe] ratios in M15 found by Sneden et al.
(\citeyear{1997AJ....114.1964S}, included in the Pritzl compilation) are
$\sim$0.4~dex higher than in our own analysis of M15, but using Sneden's
Ti lines, log~gfs and EWs (in the stars we have in common), our analysis
yields the same value as Sneden's. We also notice a small systematic
difference between the ratio of Ti~{\sc i} and Ti~{\sc ii} over iron
($\sim$0.2~dex), that could be caused by log~gfs (that could be on
different scales for Ti~{\sc i} and Ti~{\sc ii}) and/or non-LTE effects
. We therefore conclude that, based on the comparison of our Fornax
globular clusters with a fully compatible analysis of galactic globular
clusters (our analysis of M15 and three other clusters by Shetrone et
al. \citeyear{2003AJ....125..684S}), there is no difference in the Ti/Fe
ratios observed in Fornax and MW globular clusters.

\subsection{Deep mixing pattern} \label{subsec:deepmix}

Deep-mixing occurs when material processed deep inside a star finds its
way to the upper atmosphere, thus modifying the original abundance
pattern.  Proton-capture nucleosynthesis converts O, N, Ne to Na, and Mg
to Al in the H fusion layer of evolved RGB stars. This means that a
significant atmospheric depletion of O caused by deep-mixing should be
accompanied by an enhancements of Na (Langer et al.
\citeyear{1993PASP..105..301L})  and similarly an enhancement in Al
should cause observable Mg depletion (Langer \& Hoffman
\citeyear{1995PASP..107.1177L}).  Such patterns (anti-correlations of
O-Na and Mg-Al) are found in galactic globular cluster stars but not in
comparable field stars of our Galaxy (Gratton et al.
\citeyear{2004ARA&A..42..385G}), or any other (e.g., Shetrone et al.
\citeyear{2001ApJ...548..592S}).  It is assumed that this is caused by
environmental effects within a star cluster but whether it is the result
of deep-mixing within the RGB stars that are observed or the fossil
traces of self-pollution of the globular cluster during its formation
process, or a combination of the two, is not well understood.

\begin{figure}[!htb]
\begin{center}
\includegraphics[angle=0,width=0.5\textwidth]{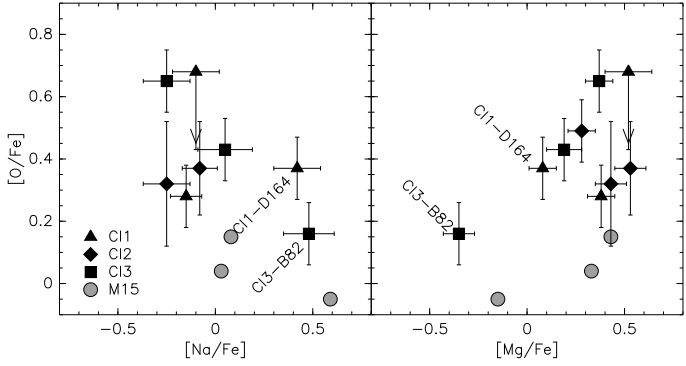}

\caption{Here we show the ``Deep-Mixing'' 
abundance anomaly.  An 
anti-correlation of Mg-Na on the left and
a correlation of O-Mg on the right. 
The symbols are the same as in 
Figure~\ref{fig:alpha_feh}.
\label{fig:anomaly}}

\end{center}
\end{figure}

Figure~\ref{fig:anomaly} shows that deep mixing patterns are not only
found in galactic globular clusters and the old clusters of the Large
Magellanic Cloud [LMC] (Hill et al. \citeyear{2000A&A...364L..19H}) but
also in clusters of much smaller dwarf spheroidal galaxies like Fornax.
The anti-correlation O-Na is visible in two (Cl3-B82 and Cl1-D164) of
the nine stars we observed in the Fornax globular clusters displaying
high Na and low O abundances (left panel), accompanied by low Mg
abundances (correlation O-Mg, right panel).  We cannot check directly
whether the Mg-Al anti-correlation also exists in these clusters, since
we did not detect Al in our Fornax dSph globular cluster spectra,
because the Al lines present in our spectral range are too weak.  Our
detection limit is about 14~m\AA{}, which translates into an upper limit
to [Al/Fe] of 1.4. Shetrone et al. (\citeyear{1996AJ....112.2639S})
found that the usual enhancement of Al ranges from 0.5 to 1.0 dex, thus
largely consistent with our upper limit.

\subsection{Iron-peak elements}

\begin{figure}[!htb]
\begin{center}
\includegraphics[angle=0,width=0.5\textwidth]{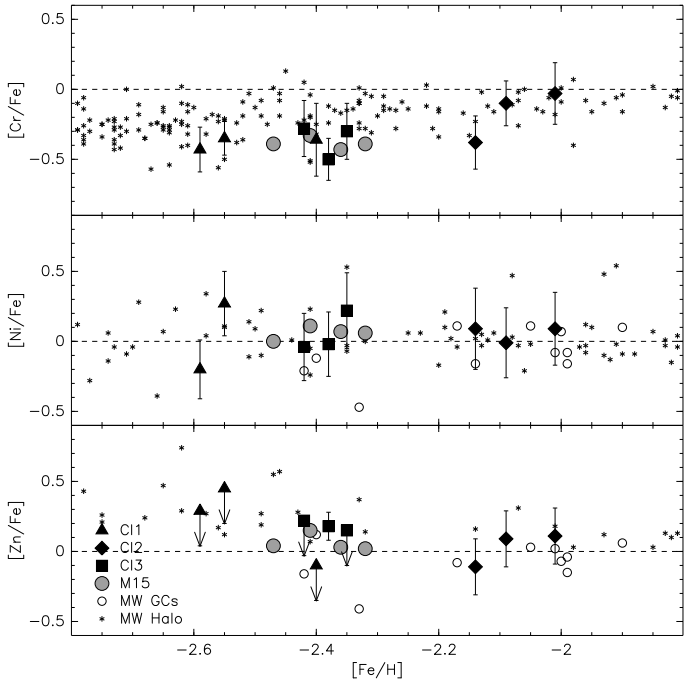}

\caption{Iron-peak elements abundances as a function of [Fe/H]. 
The symbols are the same as in Figure~\ref{fig:alpha_feh}.
\label{fig:ironp_feh}}

\end{center}
\end{figure}

The Fe-peak elements we observed in the Fornax dSph globular clusters
are Cr, Ni and Zn, and they are shown in Figure~\ref{fig:ironp_feh}.
Comparison points for Galactic globular clusters are from the
compilation of Pritzl et al. (\citeyear{2005AJ....130.2140P}), and
Galactic halo stars are from the Hamburg-ESO (HERES) survey (Barklem et
al. \citeyear{2005A&A...439..129B}) for Cr (top panel), from the
compilation of Venn et al. (\citeyear{2004AJ....128.1177V}) for Ni
(middle panel), and from Sneden et al.(\citeyear{1991A&A...246..354S})
and Barklem et al. (\citeyear{2005A&A...439..129B}) for Zn (lower
panel).

Cr is believed to be produced mainly by incomplete explosive silicon
burning (Woosley \& Weaver \citeyear{1995ApJS..101..181W}).  Despite
large error bars in our measurements of the Fornax dSph globular cluster
stars, there seems to be an increase (by $\sim$0.3 dex) of the [Cr/Fe]
ratio between the two more metal-poor clusters and the more metal-rich
Cluster 2.  Such a trend of increasing [Cr/Fe] with increasing [Fe/H]
has been observed in Galactic field stars (McWilliam et al.~
\citeyear{1995AJ....109.2757M}, Carretta et al.~
\citeyear{2002AJ....124..481C}), leading to a similar $\sim$0.3 dex
increase, but over a much wider metallicity range ($-$3.5 to $-$2.).
Newer, high quality observations by Cayrel et al.
(\citeyear{2004A&A...416.1117C}) of Galactic halo stars further reduced
the observed slope of increasing [Cr/Fe] with increasing metallicity to
$\sim$0.15 dex over a [Fe/H] range from $-$2.5 to $-$4 dex, with an
extremely small intrinsic scatter ($\sigma$ = 0.05 dex).  The higher
[Cr/Fe] abundance observed in Cluster 2 therefore seems unlikely, and is
probably caused by our observational errors.

Ni is believed to be produced in complete explosive silicon burning.
We don't expect any relation between [Ni/Fe] as a function of
[Fe/H], based on what we see in the MW.  Even at this low metallicity,
the relation is flat with a value close to zero, within the error bars,
as we can see in Figure~\ref{fig:ironp_feh}.  This is consistent with
the majority of Galactic globular clusters, open clusters and halo stars
(Sneden et al. \citeyear{2004oee..symp..172S}).  So yet again, the
Fornax dSph globular clusters are similar to the normal Galactic
globular clusters.

Zn has the same origin as Ni, but it has been suggested (Heger \&
Woosley \citeyear{2002ApJ...567..532H}) that it could also be formed by
neutron capture, and be either an \rpr\  or an \spr\  element.  Our
results, more than half of which are upper limits, are consistent with
Galactic values.

\subsection{Heavy elements}

\begin{figure}
\begin{center}
\includegraphics[angle=0,width=0.5\textwidth]{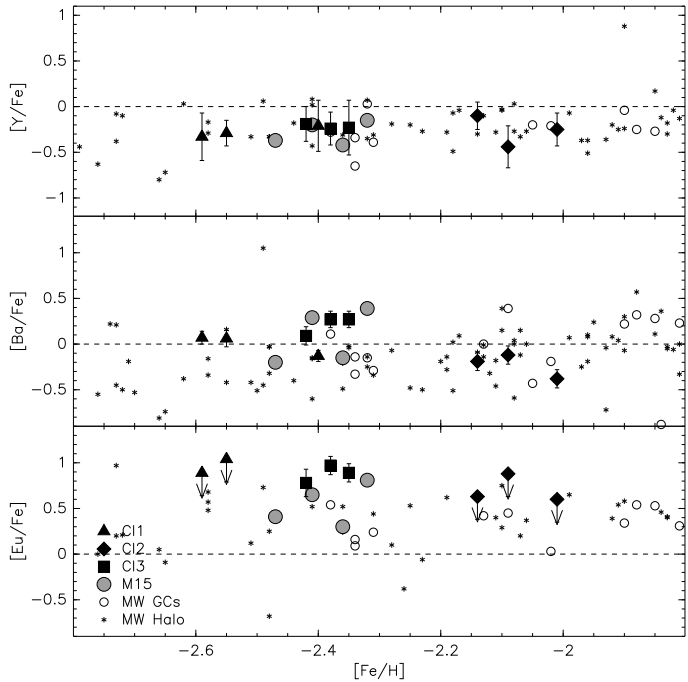}

\caption{Heavy elements abundances as a function of [Fe/H]. 
The symbols are the same as in Figure~\ref{fig:alpha_feh}. 
\label{fig:heavy_feh}}

\end{center}
\end{figure}

The heavy elements Y, Ba and Eu in the Fornax dSph globular clusters are
plotted in Figure~\ref{fig:heavy_feh}. [Y/Fe] appears to be consistent
with what is observed in Galactic globular clusters.  However, Cluster 1
and 3 (the two most metal-poor) appear to have higher [Ba/Fe] than
average for Galactic globular clusters. As shown in
Figure~\ref{fig:synthesis}), europium is measured from a
single weak line, and could only be detected in Cluster~3 (all other
Fornax points in this plot are upper limits), in which [Eu/Fe] is
particularly high, above the typical range for Galactic globular
clusters. 

\begin{figure}
\begin{center}
\includegraphics[angle=0,width=0.5\textwidth]{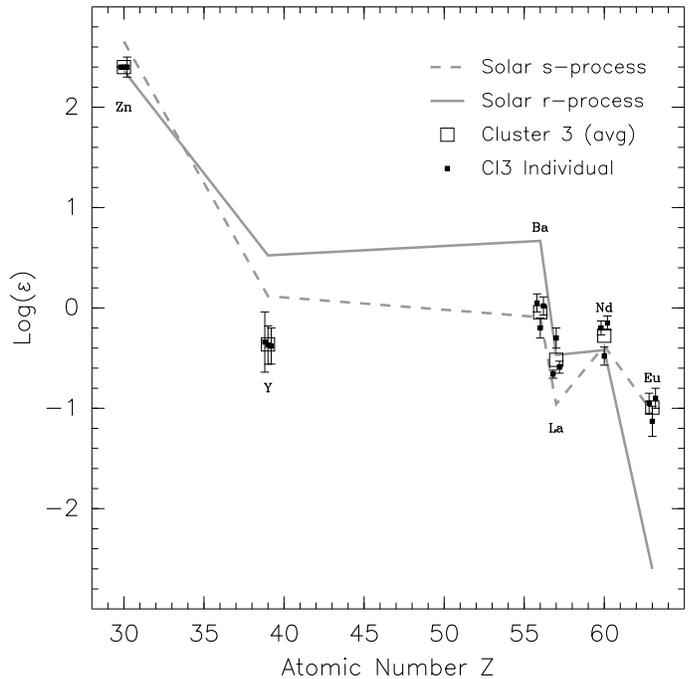}
\caption{The relative contributions of the $r-$ and $s-$ processes for
the 
heavy elements in Cluster~3 (filled circle).  The solar $r-$ and $s-$
process
abundances, traced by a dotted and a full line respectively, are taken
from Burris et al. (2000). They are shifted by the difference between
Cluster 3 and the solar system abundance for Eu ($-$1.55 dex).
\label{fig:rpr-spr}}

\end{center}
\end{figure}

Ba and Y are neutron-capture elements which are, in the solar system,
dominated by the \spr, a process due to low to intermediate-mass
Asymptotic Giant Branch (AGB) stars, with only a minor contribution from
the \rpr.  Eu on the other hand is almost entirely dominated by the
\rpr, which requires more extreme neutron fluxes, such as SN II
explosions (McWilliam~\citeyear{1997ARA&A..35..503M}) associated with
massive stars.  In the Milky Way, with decreasing metallicities the
\spr~ contribution gradually decreases (consistent with the timescale of
AGB evolution) so that below $\sim -2.5$dex, both in field and globular
clusters stars, all heavy elements are dominated exclusively by the
\rpr. (Johnson et al. \citeyear{2001ApJ...554..888J}, James et al.
\citeyear{2004A&A...427..825J}, Barklem et
al.~\citeyear{2005A&A...439..129B}).  In Cluster~3, we detect, not only
Eu, but also other heavy elements represented by weak lines preventing
detection in the other clusters: Nd and La.  In Figure
\ref{fig:rpr-spr}, we compare Cluster 3 log ($\epsilon$)\footnote{The
scale used for log ($\epsilon$) is the standard astronomical scale
($\mathrm{log}_{10}(N_{\mathrm{el}}/N_{{\mathrm H}}) +12$).} values to
the solar system $r-$ and $s-$ process abundances (Burris et
al.~\citeyear{2000ApJ...544..302B}). The solar system elemental
abundances have been shifted by the difference between the mean values
of Eu for Cluster~3 and the solar system abundance distribution ($-$1.55
dex).  Clearly, the abundances of most elements in the Fornax globular
clusters match the solar system \rpr~ pattern within the observational
uncertainties (with the exception of La which seems to be matched by
neither the $r-$ nor the $s-$ process patterns).  Cluster 3 stars are
obviously very close to the \rpr~ expectations, confirming that,
similarly to the most metal-poor globular clusters in the galactic Halo
(M15, M92, M68), cluster 3 is also dominated by the \rpr.  This is also
confirmed by the [Ba/Eu] ratio observed in the three Cluster~3 stars
[Ba/Eu] = $-$0.62, $-$0.69, $-$0.7 ($\pm$0.20), very close to the
$-0.69$ for the \rpr\  component in the solar system (as compared to
+1.15 for the \spr, Arlandini et al.  \citeyear{1999ApJ...525..886A}).
The upper limits for Eu in Clusters 1 and 2, although not decisive, are
also compatible with a pure \rpr\  enrichment ([Ba/Eu]$> -$1.00 to
$-$0.82).  This result indicates that, in Fornax dSph as in our Galaxy,
heavier neutron capture elements in the lowest metallicity stars have
only very weak \spr\  contribution. Or in other words, that heavy
elements in Fornax dSph globular clusters, as in M15, are formed
principally through the \rpr.

The high neutron-capture element content of Cluster~3, that we attribute
to the \rpr, is similar for all three stars, and above the upper edge of
the range of values traditionally covered by the Galactic globular
clusters (Pritzl et al. \citeyear{2005AJ....130.2140P}).  $R$-process
enrichments of this order or even higher are found in Galactic halo
field stars (Barklem et al.~\citeyear{2005A&A...439..129B}), but as far
as Galactic globular clusters are concerned, the only case known to date
is M15.  Sneden et al.  (\citeyear{1997AJ....114.1964S},
\citeyear{2000AJ....120.1351S}) have established that M15 has a stellar
bi-modality with one group being strongly overabundant in [Eu/Fe] (and
[Ba/Fe]) compared to the other.  Our observations of three stars in
Cluster~3 do not provide sufficient statistics to determine if this
cluster also has a bimodality (with our 3 stars by chance happening to
belong to the high Eu group) or if all stars in Cluster~3 are Eu-rich.

Finally, despite the dispersion in Ba that seems to exist among the
three Fornax globular clusters (Cluster~3 being the most Ba and Eu
rich), Y is very similar from cluster to cluster, and comparable to the
Galactic abundances of this element (globular clusters and field stars).
This also leads Cluster~3 to have a Ba/Y ratio higher than in the two
other clusters ([Ba/Y]=+0.43 compared to +0.0 in Cluster~2, and
marginally higher than the +0.28 dex observed in Cluster~1).
Interestingly, the [Ba/Y] observed in the three Fornax clusters are yet
again very similar to that of the Galactic globular clusters and halo
field stars, whereas the (on average more metal-rich) field stars in
dwarf spheroidal galaxies have been shown to display systematically
lower [Ba/Y] than their galactic counterparts (Venn et al.
\citeyear{2004AJ....128.1177V}).

\section{Conclusions}

\begin{figure}
\begin{center}
\includegraphics[angle=0,width=0.5\textwidth]{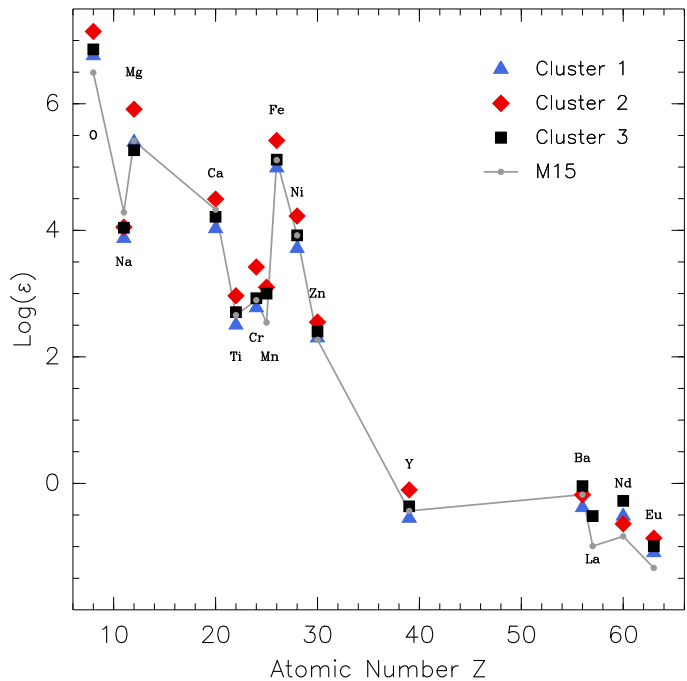}

\caption{The cluster mean elemental abundances
of the three Fornax dSph globular clusters and M15. Each individual
stellar abundance has being weighted by its error. Cluster~1 is
identified by a filled square, Cluster~2 by a star, Cluster~3 by a
cross and M15 by a triangle. 
\label{fig:loge_Z}}

\end{center}
\end{figure}

We have compared the properties of the globular clusters belonging to
the Fornax dSph with those of the Milky Way with unprecedented accuracy.
The Fornax dSph contains clusters with a range of properties such as
metallicity, central concentration and Horizontal Branch structure. For
the first time detailed chemical abundances have been derived for a
sample of stars in a globular cluster {\it system} in an external
galaxy, apart from the Magellanic Clouds.  Despite their very different
mass, morphology and global star formation history, the Fornax dSph and
the Milky Way appear to have experienced the same very early enrichment
conditions and in particular similar nucleosynthesis.  This is
summarised in Figure~\ref{fig:loge_Z}, where the mean elemental
abundances, each being weighted by its error, of the three Fornax
globular clusters and M15 are compared.  The abundance patterns of the
individual stars in Milky Way globular clusters and Fornax globular
clusters match each other almost perfectly. We find that the
star-to-star abundance dispersion in the Fornax clusters is modest and
compatible with similar observations of Galactic globular clusters.

We have definitively established that the Fornax globular Clusters 1, 2
and 3 are very metal-poor, slightly poorer than previous estimates, with
respectively \avgfeh=$ -2.5$, $-2.1$ and $-2.4$.  Part of the
discrepancy with previous studies is explained by the different
reference calibrations used.  Cluster~1 is now the most metal-poor
globular cluster known, however the difference between Cluster~1 and M92
or M15 in the Milky Way is small. There seems to be universal lower
limit to the metallicity at which star clusters form, which is higher
than that of field stars in the halo our Galaxy, where significant
numbers of stars are found with [Fe/H] $< -4$.  It is also clear, that
as in our Galaxy, the ratio of the number of globular cluster to the
number of field stars strongly decreases with rising metallicity
(Harris \& Harris~\citeyear{2002AJ....123.3108H}).

Clusters 1, 2 and 3 were clearly formed promptly and early in the
history of Fornax dSph, alike the Milky Way globular clusters. They are
over abundant in $\alpha$-elements (O, Mg, Ca) at a similar level to
Galactic clusters at identical [Fe/H], and the heavy element abundances
(Y, Ba, Eu) in the 3 clusters are compatible with dominant \rpr~
enrichment. Finally, the Fe-peak elements are also very similar to
Galactic globular cluster values, with [Ni/Fe] being unambiguously solar
in all three clusters and Zn and Cr are also compatible with Galactic
values.

The analogy between Galactic and Fornax dSph holds even in the rare
cases and anomalies: (i) Eu is extremely overabundant in Cluster~3
stars. The only Galactic counterpart known to date is M15. (ii) Cl1-D164
and Cl3-B82 show low O and Mg associated with a high Na abundance, thus
establishing an O-Na anti-correlation and O-Mg correlation.  This is the
same deep-mixing pattern observed Galactic star clusters, and old LMC
clusters.

The effort towards a comprehensive description of the formation and
evolution of the Fornax dSph will soon benefit from the analysis of
VLT/FLAMES high resolution spectra of a hundred field stars (Letarte et
al., in preparation). It will then be possible to describe the chemical
enrichment and nucleosynthetic processes dominant for the field star
population compared to that found in the globular clusters, and to see
when and if the similarities in enrichment patterns with our Galaxy end.

\begin{acknowledgements}
We gratefully acknowledge Carlo Emanuele Corsi and Roberto Buonanno for
providing their photometry for our target selection.
ET gratefully acknowledges support from a fellowship of the
Royal Netherlands Academy of Arts and Sciences.
BL is funded by a grant from the Netherlands Organisation 
for Scientific Research (NWO).
PJ and GM gratefully acknowledge support from the 
Swiss National Science Foundation (SNSF).
\end{acknowledgements}

\bibliographystyle{aa}
\bibliography{bruno}

\begin{thebibliography}{46}
\expandafter\ifx\csname natexlab\endcsname\relax\def\natexlab#1{#1}\fi

\bibitem[{{Alonso} {et~al.}(2001){Alonso}, {Arribas}, \&
  {Mart{\'{\i}}nez-Roger}}]{2001A&A...376.1039A}
{Alonso}, A., {Arribas}, S., \& {Mart{\'{\i}}nez-Roger}, C. 2001, \aap, 376,
  1039

\bibitem[{{Arlandini} {et~al.}(1999){Arlandini}, {K{\"a}ppeler}, {Wisshak},
  {Gallino}, {Lugaro}, {Busso}, \& {Straniero}}]{1999ApJ...525..886A}
{Arlandini}, C., {K{\"a}ppeler}, F., {Wisshak}, K., {et~al.} 1999, \apj, 525,
  886

\bibitem[{{Barklem} {et~al.}(2005){Barklem}, {Christlieb}, {Beers}, {Hill},
  {Bessell}, {Holmberg}, {Marsteller}, {Rossi}, {Zickgraf}, \&
  {Reimers}}]{2005A&A...439..129B}
{Barklem}, P.~S., {Christlieb}, N., {Beers}, T.~C., {et~al.} 2005, \aap, 439,
  129

\bibitem[{{Buonanno} {et~al.}(1999){Buonanno}, {Corsi}, {Castellani},
  {Marconi}, {Fusi Pecci}, \& {Zinn}}]{1999AJ....118.1671B}
{Buonanno}, R., {Corsi}, C.~E., {Castellani}, M., {et~al.} 1999, \aj, 118, 1671

\bibitem[{{Buonanno} {et~al.}(1985){Buonanno}, {Corsi}, {Fusi Pecci}, {Hardy},
  \& {Zinn}}]{1985A&A...152...65B}
{Buonanno}, R., {Corsi}, C.~E., {Fusi Pecci}, F., {Hardy}, E., \& {Zinn}, R.
  1985, \aap, 152, 65

\bibitem[{{Buonanno} {et~al.}(1998){Buonanno}, {Corsi}, {Zinn}, {Fusi Pecci},
  {Hardy}, \& {Suntzeff}}]{1998ApJ...501L..33B}
{Buonanno}, R., {Corsi}, C.~E., {Zinn}, R., {et~al.} 1998, \apjl, 501, L33+

\bibitem[{{Burris} {et~al.}(2000){Burris}, {Pilachowski}, {Armandroff},
  {Sneden}, {Cowan}, \& {Roe}}]{2000ApJ...544..302B}
{Burris}, D.~L., {Pilachowski}, C.~A., {Armandroff}, T.~E., {et~al.} 2000,
  \apj, 544, 302

\bibitem[{{Carretta} {et~al.}(2002){Carretta}, {Gratton}, {Cohen}, {Beers}, \&
  {Christlieb}}]{2002AJ....124..481C}
{Carretta}, E., {Gratton}, R., {Cohen}, J.~G., {Beers}, T.~C., \& {Christlieb},
  N. 2002, \aj, 124, 481

\bibitem[{{Cayrel} {et~al.}(2004){Cayrel}, {Depagne}, {Spite}, {Hill}, {Spite},
  {Fran{\c c}ois}, {Plez}, {Beers}, {Primas}, {Andersen}, {Barbuy},
  {Bonifacio}, {Molaro}, \& {Nordstr{\" o}m}}]{2004A&A...416.1117C}
{Cayrel}, R., {Depagne}, E., {Spite}, M., {et~al.} 2004, \aap, 416, 1117

\bibitem[{{Dekker} {et~al.}(2000){Dekker}, {D'Odorico}, {Kaufer}, {Delabre}, \&
  {Kotzlowski}}]{2000SPIE.4008..534D}
{Dekker}, H., {D'Odorico}, S., {Kaufer}, A., {Delabre}, B., \& {Kotzlowski}, H.
  2000, in Proc. SPIE Vol. 4008, p. 534-545, Optical and IR Telescope
  Instrumentation and Detectors, Masanori Iye; Alan F. Moorwood; Eds., 534--545

\bibitem[{{Demers} {et~al.}(1990){Demers}, {Grondin}, \&
  {Kunkel}}]{1990PASP..102..632D}
{Demers}, S., {Grondin}, L., \& {Kunkel}, W.~E. 1990, \pasp, 102, 632

\bibitem[{{Gratton} {et~al.}(2004){Gratton}, {Sneden}, \&
  {Carretta}}]{2004ARA&A..42..385G}
{Gratton}, R., {Sneden}, C., \& {Carretta}, E. 2004, \araa, 42, 385

\bibitem[{{Harris} \& {Harris}(2002)}]{2002AJ....123.3108H}
{Harris}, W.~E. \& {Harris}, G.~L.~H. 2002, \aj, 123, 3108

\bibitem[{{Heger} \& {Woosley}(2002)}]{2002ApJ...567..532H}
{Heger}, A. \& {Woosley}, S.~E. 2002, \apj, 567, 532

\bibitem[{{Hill} {et~al.}(2000){Hill}, {Fran{\c c}ois}, {Spite}, {Primas}, \&
  {Spite}}]{2000A&A...364L..19H}
{Hill}, V., {Fran{\c c}ois}, P., {Spite}, M., {Primas}, F., \& {Spite}, F.
  2000, \aap, 364, L19

\bibitem[{{Hodge}(1961)}]{1961PASP...73Q.328H}
{Hodge}, P.~W. 1961, \pasp, 73, 328

\bibitem[{{James} {et~al.}(2004){James}, {Fran{\c c}ois}, {Bonifacio},
  {Carretta}, {Gratton}, \& {Spite}}]{2004A&A...427..825J}
{James}, G., {Fran{\c c}ois}, P., {Bonifacio}, P., {et~al.} 2004, \aap, 427,
  825

\bibitem[{{Johnson} \& {Bolte}(2001)}]{2001ApJ...554..888J}
{Johnson}, J.~A. \& {Bolte}, M. 2001, \apj, 554, 888

\bibitem[{{Jorgensen} \& {Jimenez}(1997)}]{1997A&A...317...54J}
{Jorgensen}, U.~G. \& {Jimenez}, R. 1997, \aap, 317, 54

\bibitem[{{Kustner}(1921)}]{1921VeBon..15....1K}
{Kustner}, F. 1921, Veroeffentlichungen des Astronomisches Institute der
  Universitaet Bonn, 15, 1

\bibitem[{{Langer} {et~al.}(1993){Langer}, {Hoffman}, \&
  {Sneden}}]{1993PASP..105..301L}
{Langer}, G.~E., {Hoffman}, R., \& {Sneden}, C. 1993, \pasp, 105, 301

\bibitem[{{Langer} \& {Hoffman}(1995)}]{1995PASP..107.1177L}
{Langer}, G.~E. \& {Hoffman}, R.~D. 1995, \pasp, 107, 1177

\bibitem[{{Lotz} {et~al.}(2004){Lotz}, {Miller}, \&
  {Ferguson}}]{2004ApJ...613..262L}
{Lotz}, J.~M., {Miller}, B.~W., \& {Ferguson}, H.~C. 2004, \apj, 613, 262

\bibitem[{{McWilliam}(1997)}]{1997ARA&A..35..503M}
{McWilliam}, A. 1997, \araa, 35, 503

\bibitem[{{McWilliam} {et~al.}(1995){McWilliam}, {Preston}, {Sneden}, \&
  {Searle}}]{1995AJ....109.2757M}
{McWilliam}, A., {Preston}, G.~W., {Sneden}, C., \& {Searle}, L. 1995, \aj,
  109, 2757

\bibitem[{{Pritzl} {et~al.}(2005){Pritzl}, {Venn}, \&
  {Irwin}}]{2005AJ....130.2140P}
{Pritzl}, B.~J., {Venn}, K.~A., \& {Irwin}, M. 2005, \aj, 130, 2140

\bibitem[{{Sandage}(1970)}]{1970ApJ...162..841S}
{Sandage}, A. 1970, \apj, 162, 841

\bibitem[{{Saviane} {et~al.}(2000){Saviane}, {Held}, \&
  {Bertelli}}]{2000A&A...355...56S}
{Saviane}, I., {Held}, E.~V., \& {Bertelli}, G. 2000, \aap, 355, 56

\bibitem[{{Seth} {et~al.}(2004){Seth}, {Olsen}, {Miller}, {Lotz}, \&
  {Telford}}]{2004AJ....127..798S}
{Seth}, A., {Olsen}, K., {Miller}, B., {Lotz}, J., \& {Telford}, R. 2004, \aj,
  127, 798

\bibitem[{{Shapley}(1938)}]{1938NAT...142...715}
{Shapley}, H. 1938, \nat, 142, 715

\bibitem[{{Shetrone} {et~al.}(2003){Shetrone}, {Venn}, {Tolstoy}, {Primas},
  {Hill}, \& {Kaufer}}]{2003AJ....125..684S}
{Shetrone}, M., {Venn}, K.~A., {Tolstoy}, E., {et~al.} 2003, \aj, 125, 684

\bibitem[{{Shetrone}(1996{\natexlab{a}})}]{1996AJ....112.1517S}
{Shetrone}, M.~D. 1996{\natexlab{a}}, \aj, 112, 1517

\bibitem[{{Shetrone}(1996{\natexlab{b}})}]{1996AJ....112.2639S}
{Shetrone}, M.~D. 1996{\natexlab{b}}, \aj, 112, 2639

\bibitem[{{Shetrone} {et~al.}(2001){Shetrone}, {C{\^ o}t{\' e}}, \&
  {Sargent}}]{2001ApJ...548..592S}
{Shetrone}, M.~D., {C{\^ o}t{\' e}}, P., \& {Sargent}, W.~L.~W. 2001, \apj,
  548, 592

\bibitem[{{Sneden} {et~al.}(1991){Sneden}, {Gratton}, \&
  {Crocker}}]{1991A&A...246..354S}
{Sneden}, C., {Gratton}, R.~G., \& {Crocker}, D.~A. 1991, \aap, 246, 354

\bibitem[{{Sneden} {et~al.}(2004){Sneden}, {Ivans}, \&
  {Fulbright}}]{2004oee..symp..172S}
{Sneden}, C., {Ivans}, I.~I., \& {Fulbright}, J.~P. 2004, in Origin and
  Evolution of the Elements, 172

\bibitem[{{Sneden} {et~al.}(1997){Sneden}, {Kraft}, {Shetrone}, {Smith},
  {Langer}, \& {Prosser}}]{1997AJ....114.1964S}
{Sneden}, C., {Kraft}, R.~P., {Shetrone}, M.~D., {et~al.} 1997, \aj, 114, 1964

\bibitem[{{Sneden} {et~al.}(2000){Sneden}, {Pilachowski}, \&
  {Kraft}}]{2000AJ....120.1351S}
{Sneden}, C., {Pilachowski}, C.~A., \& {Kraft}, R.~P. 2000, \aj, 120, 1351

\bibitem[{{Spite}(1967)}]{1967AnAp...30..211S}
{Spite}, M. 1967, Annales d'Astrophysique, 30, 211

\bibitem[{{Stetson} {et~al.}(1998){Stetson}, {Hesser}, \&
  {Smecker-Hane}}]{1998PASP..110..533S}
{Stetson}, P.~B., {Hesser}, J.~E., \& {Smecker-Hane}, T.~A. 1998, \pasp, 110,
  533

\bibitem[{{Strader} {et~al.}(2003){Strader}, {Brodie}, {Forbes}, {Beasley}, \&
  {Huchra}}]{2003AJ....125.1291S}
{Strader}, J., {Brodie}, J.~P., {Forbes}, D.~A., {Beasley}, M.~A., \& {Huchra},
  J.~P. 2003, \aj, 125, 1291

\bibitem[{{Tinsley}(1979)}]{1979ApJ...229.1046T}
{Tinsley}, B.~M. 1979, \apj, 229, 1046

\bibitem[{{van den Bergh}(2005)}]{astro-ph/0509811}
{van den Bergh}, S. 2005, astro-ph/0509811

\bibitem[{{Venn} {et~al.}(2004){Venn}, {Irwin}, {Shetrone}, {Tout}, {Hill}, \&
  {Tolstoy}}]{2004AJ....128.1177V}
{Venn}, K.~A., {Irwin}, M., {Shetrone}, M.~D., {et~al.} 2004, \aj, 128, 1177

\bibitem[{{Woosley} \& {Weaver}(1995)}]{1995ApJS..101..181W}
{Woosley}, S.~E. \& {Weaver}, T.~A. 1995, \apjs, 101, 181

\bibitem[{{Zinn} \& {West}(1984)}]{1984ApJS...55...45Z}
{Zinn}, R. \& {West}, M.~J. 1984, \apjs, 55, 45

\end{thebibliography}

\appendix 
\renewcommand{\thetable}{A.\arabic{table}} 


\begin{singlespace}
\clearpage
\onecolumn
\begin{landscape}
\begin{center}
\small
\begin{longtable}{clrrcrrrrrrrrrrrrr}
\caption{Line parameters and equivalent widths for the Fornax globular clusters and M15.
When there is a * in the S column, it indicate that a syntetic spectra 
was used for the abondance determination. HFS indicate a line with 
hyperfine splitting, so no individual EW measurment for that line is available. 
\label{tab:LL_EW}}\\
\hline \hline
$\lambda$&     El&    E.P.&  log gf&  S& Cl1-D56& Cl1-D68& Cl1-D164&   Cl2-B71&   Cl2-B77& Cl2-B226&     Cl3-B59& Cl3-B61&  Cl3-B82&  M15S1&  M15S3&  M15S6&  M15S7\\
\hline
\endhead
\hline \multicolumn{18}{r}{\emph{Continued on next page}}
\endfoot
\hline
\endlastfoot
4934.12&   Ba~{\sc ii}&    0.00&  -0.703&  *&  HFS  &   HFS  &    HFS  &     HFS  &       ...&      ...&    HFS  &       ...&       ...&    HFS  &    HFS  &   HFS  &    HFS  \\
5853.69&   Ba~{\sc ii}&    0.60&  -1.010&  *&  HFS  &   HFS  &    HFS  &     HFS  &     HFS  &    HFS  &      ...&     HFS  &       ...&    HFS  &    HFS  &   HFS  &    HFS  \\
6141.73&   Ba~{\sc ii}&    0.70&  -0.077&  *&  HFS  &   HFS  &    HFS  &     HFS  &     HFS  &    HFS  &    HFS  &     HFS  &     HFS  &    HFS  &    HFS  &   HFS  &    HFS  \\
6496.91&   Ba~{\sc ii}&    0.60&  -0.380&  *&  103.7&   136.8&    123.6&     121.5&     115.0&    138.2&    144.8&     123.1&     147.2&    124.8&    136.2&   149.2&    118.4\\
6102.73&    Ca~{\sc i}&    1.88&  -0.790&   &   50.4&    54.5&     50.8&      91.9&      97.3&    116.0&     60.2&      63.4&      72.4&     77.7&     74.8&    79.9&     71.5\\
6122.23&    Ca~{\sc i}&    1.89&  -0.320&   &   85.0&    93.5&     86.5&     129.0&     119.2&    160.1&    107.0&     104.5&     103.8&    116.1&    110.6&   114.2&    105.9\\
6161.30&    Ca~{\sc i}&    2.52&  -1.270&   &    ...&     ...&      ...&       ...&       ...&     26.3&      ...&       ...&       ...&     11.8&     14.0&     9.7&      9.0\\
6166.44&    Ca~{\sc i}&    2.52&  -1.140&   &    ...&     ...&      ...&      27.7&       ...&     32.8&      ...&       ...&       ...&     17.6&     16.7&     ...&     16.9\\
6169.04&    Ca~{\sc i}&    2.52&  -0.800&   &   30.2&    14.6&     12.7&      33.3&      41.7&     51.5&      ...&      21.4&      33.6&     28.4&     28.4&    32.0&     24.7\\
6169.56&    Ca~{\sc i}&    2.52&  -0.480&   &    ...&    24.1&     17.9&      41.1&      45.7&     58.8&      ...&      41.0&       ...&     43.4&     39.2&    44.2&     38.3\\
6439.08&    Ca~{\sc i}&    2.52&   0.390&   &   77.5&    69.2&     72.4&     103.1&     114.0&    139.8&     84.5&      88.2&      97.1&     99.7&     95.7&    99.4&     95.9\\
6455.60&    Ca~{\sc i}&    2.52&  -1.290&   &    ...&     ...&      ...&       ...&       ...&     30.3&     26.0&       ...&       ...&     13.1&     10.5&     9.5&     10.0\\
6499.65&    Ca~{\sc i}&    2.52&  -0.820&   &   24.1&    22.1&     11.2&      31.3&      35.1&     58.9&      ...&      20.0&      25.6&     26.8&     23.4&    26.4&     23.6\\
5206.04&    Cr~{\sc i}&    0.94&   0.019&   &  100.0&   124.2&    109.0&     126.4&     170.5&    214.4&    136.3&     129.2&     129.3&    126.9&    115.7&   121.3&    112.2\\
5409.80&    Cr~{\sc i}&    1.03&  -0.720&   &   66.3&    74.2&     74.8&      95.5&     114.1&    148.9&     89.3&      85.3&      86.3&     88.1&     78.8&    84.5&     74.0\\
6645.13&   Eu~{\sc ii}&    1.37&   0.200&  *&    ...&   HFS  &    HFS  &     HFS  &     HFS  &    HFS  &    HFS  &     HFS  &     HFS  &    HFS  &    HFS  &   HFS  &    HFS  \\
4966.10&    Fe~{\sc i}&    3.33&  -0.890&   &   51.3&    54.0&     39.8&      68.2&      84.0&     85.1&     53.5&      42.6&      57.8&     63.3&     60.4&    64.2&     51.9\\
5006.12&    Fe~{\sc i}&    2.83&  -0.628&   &   82.0&    91.4&    108.2&     117.1&     112.4&    140.0&    119.5&     109.7&     113.7&    107.0&    100.9&   106.7&     97.3\\
5079.75&    Fe~{\sc i}&    0.99&  -3.240&   &    ...&   127.1&    119.6&     141.9&     136.6&    193.0&    133.1&     116.6&     130.0&    118.6&    109.7&   115.2&    101.7\\
5083.35&    Fe~{\sc i}&    0.96&  -2.862&   &  100.3&   124.7&    122.3&     138.8&     135.0&    181.7&    126.3&     114.5&     134.0&    130.5&    120.6&   122.2&    114.8\\
5150.85&    Fe~{\sc i}&    0.99&  -3.030&   &   84.7&   114.4&    112.7&     117.4&     129.9&    178.3&    150.6&     114.2&     128.0&    122.4&    106.3&   111.9&    100.6\\
5151.92&    Fe~{\sc i}&    1.01&  -3.326&   &   69.7&   115.4&     90.3&     104.4&     118.9&    155.9&    114.1&     120.3&     143.8&    111.5&      ...&   102.6&     91.4\\
5162.29&    Fe~{\sc i}&    4.18&   0.020&   &   64.5&    32.0&     44.7&      76.3&      67.2&     90.2&     50.0&      62.3&      46.7&     49.9&     44.1&    49.9&     43.0\\
5166.28&    Fe~{\sc i}&    0.00&  -4.200&   &  110.5&     ...&    130.8&     140.5&     180.3&    198.8&    153.3&     159.4&     146.4&    144.6&    128.0&   131.9&    121.0\\
5171.61&    Fe~{\sc i}&    1.48&  -1.751&   &  126.3&   133.7&    126.4&     132.6&     149.0&    178.1&    141.7&     142.0&     142.3&    145.8&    133.8&   142.0&    133.7\\
5192.34&    Fe~{\sc i}&    3.00&  -0.520&   &   98.0&    89.1&     79.0&     104.7&     124.7&    134.1&    122.7&      82.0&     110.5&    102.2&     96.8&    97.4&     96.0\\
5196.08&    Fe~{\sc i}&    4.26&  -0.450&   &    ...&     ...&      ...&      41.8&       9.3&     33.2&     18.7&      23.4&       ...&     11.8&      9.4&     8.8&      9.1\\
5215.19&    Fe~{\sc i}&    3.27&  -0.930&   &   41.4&    35.8&     43.9&      76.4&      81.9&     86.7&     58.2&      47.0&      47.3&     53.8&     52.7&    56.0&     47.6\\
5216.28&    Fe~{\sc i}&    1.61&  -2.102&   &   97.6&   104.9&    107.0&     115.1&     128.2&    165.4&    130.4&     119.2&     135.9&    123.1&    110.6&   118.4&    108.5\\
5217.30&    Fe~{\sc i}&    3.21&  -1.270&   &   36.0&     ...&     45.4&      63.4&       ...&     93.3&     44.5&      43.2&       ...&     50.5&     48.7&    49.9&      ...\\
5232.95&    Fe~{\sc i}&    2.94&  -0.067&   &  122.3&   122.8&    104.6&     130.9&     135.9&    166.2&    125.1&     113.3&     123.2&    124.5&    118.8&   122.1&    112.3\\
5250.21&    Fe~{\sc i}&    0.12&  -4.700&   &   59.0&    78.4&     78.8&      95.0&      99.2&    145.1&    102.3&      87.7&      95.2&     92.5&     76.5&    83.3&     67.9\\
5307.37&    Fe~{\sc i}&    1.61&  -2.812&   &   58.2&    76.3&     70.9&      91.7&      92.4&    129.1&     74.3&      74.1&      83.6&     77.8&     68.2&    73.3&     65.8\\
5324.19&    Fe~{\sc i}&    3.21&  -0.100&   &   86.4&    84.0&    100.6&     127.7&     114.8&    146.2&    106.4&     100.7&     105.9&    105.7&     99.6&   104.3&     84.3\\
5339.93&    Fe~{\sc i}&    3.27&  -0.680&   &   65.8&    59.3&     52.1&     111.3&      99.2&    123.6&     86.6&      71.5&      92.7&     72.7&     66.7&    72.1&     59.0\\
5364.86&    Fe~{\sc i}&    4.45&   0.220&   &    ...&     ...&     32.8&      52.1&      92.8&     72.2&     53.6&      45.5&      40.7&     40.5&     37.6&    40.5&     32.3\\
5367.48&    Fe~{\sc i}&    4.42&   0.550&   &   34.3&     ...&     37.7&      54.7&      56.7&     78.2&     51.5&      47.8&      42.6&     45.8&     42.8&    46.5&     38.8\\
5369.96&    Fe~{\sc i}&    4.37&   0.540&   &   64.2&     ...&     26.0&      74.3&      68.3&     88.6&     72.1&      55.3&      59.0&     49.9&     50.6&    50.5&     47.5\\
5371.50&    Fe~{\sc i}&    0.96&  -1.644&   &  205.6&   195.4&    188.1&     171.9&     196.1&    273.1&    191.1&     184.0&     201.7&    187.2&    177.4&   180.7&    171.3\\
5383.37&    Fe~{\sc i}&    4.31&   0.500&   &   49.7&    49.1&     51.4&      79.1&      59.9&    102.9&     60.3&      60.0&      64.6&     61.6&     57.7&    59.6&     50.9\\
5393.17&    Fe~{\sc i}&    3.24&  -0.920&   &   58.3&    57.2&     70.4&      80.9&     116.0&    111.7&     76.1&      70.8&      65.3&     71.6&     63.6&    69.0&     61.2\\
5397.14&    Fe~{\sc i}&    0.91&  -1.992&   &  161.7&   181.1&    172.3&     182.3&     198.0&    246.3&    171.5&     180.6&     190.1&    177.9&    163.5&   169.4&    157.8\\
5405.79&    Fe~{\sc i}&    0.99&  -1.852&   &  159.2&   190.7&    163.8&     173.6&     201.6&    230.3&    183.9&     184.4&     182.0&    175.3&    165.0&   172.6&    157.7\\
5415.19&    Fe~{\sc i}&    4.39&   0.510&   &   54.3&     ...&     50.3&      68.3&      91.9&     96.0&     48.4&      41.0&      58.6&     59.9&     55.5&    57.1&     50.4\\
5424.07&    Fe~{\sc i}&    4.32&   0.520&   &   72.8&    67.7&     57.2&      80.9&      94.1&     91.2&     78.3&      72.4&      71.0&     65.1&     63.0&    66.0&     55.5\\
5501.48&    Fe~{\sc i}&    0.96&  -3.050&   &  114.1&   132.9&    124.7&     143.4&     140.8&    174.0&    133.0&     137.9&     149.6&    135.2&    120.4&   127.1&    114.5\\
5506.79&    Fe~{\sc i}&    0.99&  -2.790&   &  111.2&   131.9&    134.3&     157.9&     149.1&    206.1&    142.3&     122.2&     150.1&    144.3&    131.0&   135.3&    127.8\\
5615.66&    Fe~{\sc i}&    3.33&   0.050&   &   84.7&    88.8&     94.9&     115.0&     124.5&    149.4&    104.0&      95.0&     111.9&    105.8&     98.5&   102.8&     95.0\\
5956.70&    Fe~{\sc i}&    0.86&  -4.570&   &   40.8&    37.5&      ...&      51.1&       ...&     99.2&     49.8&      61.8&      55.8&     49.9&     37.9&    37.0&     29.0\\
6024.05&    Fe~{\sc i}&    4.55&  -0.110&   &   18.6&    12.0&      ...&      28.8&      49.1&     54.3&     32.1&      29.1&      22.7&     25.3&     20.7&    24.6&     19.8\\
6136.62&    Fe~{\sc i}&    2.45&  -1.500&   &   98.8&    91.3&     92.3&     119.8&     109.6&    159.3&    100.0&     104.5&     122.5&    110.5&     98.9&   106.1&     94.4\\
6137.70&    Fe~{\sc i}&    2.59&  -1.366&   &   77.7&    77.1&     83.7&     109.1&     113.7&    143.9&    104.0&      88.9&     105.5&     98.8&     93.0&    99.0&     89.4\\
6157.75&    Fe~{\sc i}&    4.07&  -1.260&   &    ...&     ...&      ...&       ...&       ...&      ...&      ...&       ...&       ...&      8.8&      8.5&     9.6&      6.3\\
6173.34&    Fe~{\sc i}&    2.22&  -2.850&   &   29.8&    21.5&     21.8&      45.8&      62.2&     80.5&     34.4&      32.3&      36.8&     40.2&     32.9&    39.9&     33.0\\
6191.57&    Fe~{\sc i}&    2.43&  -1.416&   &   76.3&    92.6&     93.4&     121.7&     117.8&    140.8&    123.1&      86.8&     109.4&    109.0&    100.7&   101.4&     94.7\\
6213.43&    Fe~{\sc i}&    2.22&  -2.660&   &   34.6&    39.9&     41.0&      68.5&      59.0&     91.0&     45.2&      50.0&      47.8&     57.9&     52.1&    55.5&     45.5\\
6219.29&    Fe~{\sc i}&    2.20&  -2.438&   &   46.6&    60.3&     52.4&      87.5&      86.6&    109.0&     78.6&      64.7&      60.6&     66.6&     59.5&    68.7&     53.0\\
6229.23&    Fe~{\sc i}&    2.84&  -2.900&   &    ...&     ...&      ...&       ...&      33.0&     24.7&      ...&       ...&      16.7&      8.3&      7.7&    11.5&      7.0\\
6230.74&    Fe~{\sc i}&    2.56&  -1.276&   &   84.1&    93.2&     94.6&     115.6&      49.3&    156.8&    107.2&     105.2&     116.5&    115.2&    105.4&   112.6&    102.2\\
6232.64&    Fe~{\sc i}&    3.65&  -0.960&   &    ...&     ...&     16.5&      40.3&      45.1&     45.3&      ...&      19.0&       ...&      ...&      ...&     ...&      ...\\
6252.57&    Fe~{\sc i}&    2.40&  -1.757&   &   82.8&    82.9&     98.1&     110.2&     124.0&    146.5&    105.2&     100.7&     100.9&    101.7&     95.0&    94.3&     86.8\\
6270.23&    Fe~{\sc i}&    2.85&  -2.610&   &    ...&     ...&      ...&      16.9&      24.1&     37.5&      ...&       ...&      20.4&      ...&      ...&     ...&      ...\\
6297.80&    Fe~{\sc i}&    2.22&  -2.740&   &    ...&    57.3&     61.8&      68.8&      80.5&    113.5&      ...&      70.0&      59.9&     69.0&     65.7&     ...&      ...\\
6301.50&    Fe~{\sc i}&    3.65&  -0.720&   &   62.0&    51.3&     44.4&      68.0&      81.4&    106.5&     50.7&      53.1&      50.1&     45.2&     43.7&    44.7&     36.3\\
6302.49&    Fe~{\sc i}&    3.69&  -1.150&   &    ...&     9.2&      8.8&      23.8&      51.0&     57.1&      ...&       ...&       ...&      ...&     21.9&    24.8&      ...\\
6393.61&    Fe~{\sc i}&    2.43&  -1.630&   &   68.8&    94.3&     95.9&      86.4&     137.7&    154.7&     94.8&      72.0&     102.6&    104.9&     93.2&    99.3&     91.1\\
6421.36&    Fe~{\sc i}&    2.28&  -2.014&   &   52.9&    72.9&     71.2&      90.4&     136.0&    142.0&     91.1&      80.7&      90.5&     94.7&     83.9&    91.0&     79.8\\
6430.86&    Fe~{\sc i}&    2.18&  -1.946&   &   66.9&    87.8&     88.5&      94.7&       ...&    148.3&     98.4&      97.5&     110.6&    105.3&     92.1&    98.5&     90.3\\
6481.87&    Fe~{\sc i}&    2.27&  -2.980&   &    ...&    27.9&     23.9&      42.8&      57.0&     79.8&     25.7&      37.5&      35.0&      ...&      ...&     ...&      ...\\
6498.94&    Fe~{\sc i}&    0.96&  -4.690&   &    ...&     ...&     30.1&      38.1&      57.7&     98.7&     35.8&      37.0&      40.1&     41.0&     28.6&    34.3&     23.3\\
6518.37&    Fe~{\sc i}&    2.83&  -2.460&   &    ...&     ...&      ...&       ...&      23.8&     47.7&      ...&      19.2&      17.1&     28.7&     29.4&     ...&      ...\\
6574.23&    Fe~{\sc i}&    0.99&  -5.020&   &    ...&    11.5&      7.5&       ...&      38.2&     64.3&      ...&      19.9&      29.2&     23.2&     16.5&    20.8&     13.9\\
6593.88&    Fe~{\sc i}&    2.43&  -2.390&   &    ...&    33.6&     27.2&      69.8&      66.1&     89.1&     57.6&      42.6&      59.5&     52.9&     43.2&    53.2&     39.4\\
6609.12&    Fe~{\sc i}&    2.56&  -2.660&   &    ...&    15.1&      6.7&      37.4&      55.7&     63.6&      ...&      36.6&       ...&     24.7&     20.4&    20.8&     16.3\\
4923.92&   Fe~{\sc ii}&    2.89&  -1.320&   &  117.7&   125.4&    122.9&     130.4&     114.8&    132.4&    117.3&     115.6&     117.7&    122.4&    121.7&   124.8&    118.6\\
5197.57&   Fe~{\sc ii}&    3.23&  -2.100&   &   58.3&    43.8&     49.7&      65.6&      77.3&     78.4&     38.6&      52.9&      45.4&     54.9&     51.6&    57.5&     51.2\\
5234.63&   Fe~{\sc ii}&    3.22&  -2.118&   &   64.1&    52.6&     52.1&       ...&      69.2&     80.0&     55.1&     121.1&       ...&     59.0&     58.9&    59.7&     51.4\\
5276.00&   Fe~{\sc ii}&    3.20&  -1.950&   &   96.5&    50.1&     55.3&      87.9&      93.2&     92.4&     87.8&      68.5&      77.3&     69.7&     69.7&    68.3&     61.4\\
5284.10&   Fe~{\sc ii}&    2.89&  -3.190&   &    ...&     ...&     21.2&       ...&      37.0&     52.8&     40.7&      41.9&      40.5&     29.9&     27.5&    31.6&     24.6\\
5325.56&   Fe~{\sc ii}&    3.22&  -2.600&   &    ...&     ...&      ...&       ...&       ...&      ...&      9.1&       ...&       ...&     14.9&     14.2&    15.4&     14.2\\
5425.25&   Fe~{\sc ii}&    3.20&  -3.360&   &    ...&     ...&     22.8&      34.5&       ...&      ...&      ...&       ...&       ...&     14.1&     10.8&    13.0&      9.3\\
5534.85&   Fe~{\sc ii}&    3.24&  -2.920&   &   27.9&    27.8&     24.4&      46.8&      42.7&     43.9&     38.6&       ...&      27.8&     28.0&     27.7&    26.2&     24.0\\
5991.38&   Fe~{\sc ii}&    3.15&  -3.740&   &    ...&     ...&      ...&       ...&       ...&      ...&      ...&       ...&       ...&     14.2&      8.4&    14.8&     10.1\\
6149.25&   Fe~{\sc ii}&    3.89&  -2.720&   &    ...&     ...&      ...&       ...&      24.1&      ...&      ...&       ...&       ...&      6.5&      4.7&     8.8&      8.6\\
6238.38&   Fe~{\sc ii}&    3.89&  -2.480&   &    ...&     ...&      ...&       ...&       ...&      ...&      ...&       ...&       ...&     13.8&     12.2&    14.9&     11.8\\
6369.46&   Fe~{\sc ii}&    2.89&  -4.250&   &    ...&     ...&      ...&       ...&       ...&      ...&      ...&       ...&       ...&     10.1&      ...&     ...&      ...\\
6432.68&   Fe~{\sc ii}&    2.89&  -3.710&   &    ...&    17.5&      ...&      19.0&       ...&      ...&      ...&       ...&      20.0&     18.1&     16.7&    20.3&     18.2\\
6456.39&   Fe~{\sc ii}&    3.90&  -2.080&   &    ...&     6.8&     19.8&       ...&       ...&     32.2&      ...&       ...&       ...&     26.6&     28.0&    28.6&     26.5\\
6516.08&   Fe~{\sc ii}&    2.89&  -3.450&   &    ...&     ...&      ...&      35.9&      35.1&     40.8&      ...&       ...&      27.9&     31.6&     26.7&    26.4&     23.3\\
5301.97&   La~{\sc ii}&    0.40&  -1.140&  *&    ...&     ...&      ...&       ...&       ...&      ...&    16.0 &       ...&     22.3 &      ...&      ...&     ...&      ...\\
5303.52&   La~{\sc ii}&    0.32&  -1.350&  *&    ...&     ...&      ...&       ...&       ...&      ...&    12.9 &       ...&     13.7 &      ...&      ...&     ...&      ...\\
6320.43&   La~{\sc ii}&    0.17&  -1.562&  *&    ...&     ...&      ...&       ...&       ...&      ...&    19.3 &     15.6 &     16.9 &      ...&      ...&     ...&      ...\\
6390.46&   La~{\sc ii}&    0.32&  -1.400&  *&    ...&     ...&      ...&       ...&       ...&      ...&    18.2 &     25.8 &     23.4 &      ...&      ...&     ...&      ...\\
6774.27&   La~{\sc ii}&    0.13&  -1.708&  *&    ...&     ...&      ...&       ...&       ...&      ...&    14.2 &     9.2  &     15.2 &      ...&      ...&     ...&      ...\\
5172.70&    Mg~{\sc i}&    2.71&  -0.390&  *&  283.0&   266.1&    241.8&     312.6&     340.9&      ...&    276.4&     260.0&     204.2&    264.1&    266.4&   237.1&    247.9\\
5528.41&    Mg~{\sc i}&    4.35&  -0.357&  *&  110.2&   93.3 &    79.0 &     130.7&     141.1&    135.8&    96.0 &     102.6&     60.0 &    106.7&    101.5&   81.1 &    108.1\\
5711.09&    Mg~{\sc i}&    4.35&  -1.728&  *&  26.2 &   12.1 &    8.1  &     48.5 &     46.9 &    53.1 &    18.1 &     27.6 &     8.6  &    22.5 &    20.0 &   11.1 &    19.6 \\
6013.51&    Mn~{\sc i}&    3.07&  -0.252&  *&    ...&     ...&      ...&       ...&       ...&    25.9 &      ...&     15.0 &     15.6 &      ...&      ...&     ...&      ...\\
6021.82&    Mn~{\sc i}&    3.08&   0.035&  *&    ...&     ...&      ...&       ...&       ...&    40.9 &      ...&     25.0 &     26.2 &      ...&      ...&     ...&      ...\\
5889.97&    Na~{\sc i}&    0.00&   0.122&  *&  230.5&   219.7&    249.6&     230.4&     231.4&      ...&    242.1&     206.2&     251.6&    240.6&    296.2&   302.3&    237.7\\
5895.94&    Na~{\sc i}&    0.00&  -0.184&  *&  204.0&   185.1&    240.0&     216.1&     215.6&      ...&    217.3&     199.4&     271.0&    216.2&    257.3&   261.2&    194.3\\
6154.23&    Na~{\sc i}&    2.10&  -1.560&   &    ...&     ...&      ...&       ...&       ...&      ...&      ...&       ...&       ...&      ...&      ...&     ...&      ...\\
5249.59&   Nd~{\sc ii}&    0.98&   0.217&  *&    ...&   27.5 &      ...&       ...&       ...&    21.1 &    45.2 &     14.4 &     46.3 &     12.5&     21.8&    23.0&      8.3\\
5319.82&   Nd~{\sc ii}&    0.55&  -0.194&  *&  37.9 &   30.8 &    30.7 &       ...&       ...&    33.5 &    55.2 &     37.2 &     62.8 &      ...&     31.8&    38.0&      ...\\
5476.92&    Ni~{\sc i}&    1.83&  -0.890&   &  152.1&   124.5&     96.3&      96.5&     131.6&    158.1&    134.7&     110.2&     123.4&    115.9&    108.4&   108.8&    100.9\\
6176.82&    Ni~{\sc i}&    4.09&  -0.430&   &    ...&     ...&      8.6&      22.1&       ...&      ...&      ...&       ...&       ...&      9.4&      7.8&     9.2&      8.8\\
6177.25&    Ni~{\sc i}&    1.83&  -3.500&   &    ...&     ...&      3.4&       ...&       ...&     21.1&      ...&       ...&       ...&      9.0&      5.8&     6.2&      5.2\\
6300.31&     O~{\sc i}&    0.00&  -9.760&  *&  19.3 &   11.9 &    10.0 &     21.6 &     23.8 &    45.0 &    20.8 &     22.7 &     11.9 &    10.7 &      ...&   7.2  &    10.0 \\
4840.87&    Ti~{\sc i}&    0.90&  -0.450&   &   40.6&    23.4&     17.6&      48.8&      64.8&     69.8&     33.9&       ...&      37.9&     40.1&     36.2&    35.8&     30.2\\
4913.62&    Ti~{\sc i}&    1.87&   0.216&   &    ...&     ...&     18.5&      39.0&       ...&     41.2&     25.5&      22.0&       ...&     14.4&      5.7&    15.3&     10.9\\
5014.24&    Ti~{\sc i}&    0.81&   0.910&   &   97.2&    96.3&     95.3&     154.4&     173.8&    204.8&    122.3&     122.0&     124.6&      ...&      ...&     ...&      ...\\
5016.16&    Ti~{\sc i}&    0.85&  -0.510&   &   31.4&    26.5&      ...&      43.7&      69.2&     80.3&     43.2&       ...&      36.8&     45.7&     36.3&    41.5&     37.4\\
5064.65&    Ti~{\sc i}&    0.05&  -0.930&   &   60.9&    73.3&     63.8&      86.1&      99.7&    148.0&     88.4&      78.8&      97.6&     89.2&     75.6&    81.8&     74.6\\
5210.39&    Ti~{\sc i}&    0.05&  -0.580&   &   60.5&    76.3&     77.2&      98.1&     115.0&    164.9&    111.1&     105.7&      98.6&     94.2&     83.5&    84.3&     77.6\\
4798.53&   Ti~{\sc ii}&    1.08&  -2.670&   &   54.1&    36.3&      ...&      45.6&       ...&      ...&     54.6&      45.1&       ...&      ...&      ...&     ...&      ...\\
5129.16&   Ti~{\sc ii}&    1.89&  -1.390&   &   63.2&    54.4&     55.5&      63.4&     104.1&     96.5&     79.4&      74.1&      63.7&     74.5&     72.1&    73.1&     68.7\\
5154.07&   Ti~{\sc ii}&    1.57&  -1.520&   &   62.2&    69.6&     71.2&       ...&      78.7&     96.1&     80.2&      92.2&      81.2&     77.3&     73.2&    75.3&     66.6\\
5226.55&   Ti~{\sc ii}&    1.57&  -1.000&   &   90.2&    86.7&     97.6&     105.2&     124.6&    126.2&    109.7&      91.4&     103.6&    106.2&    101.1&   104.5&     98.2\\
5381.01&   Ti~{\sc ii}&    1.57&  -1.780&   &   54.2&    39.9&     47.2&      71.8&      81.5&     97.1&     78.0&      67.5&      65.4&     64.2&     56.8&    62.5&     54.5\\
5418.77&   Ti~{\sc ii}&    1.58&  -2.110&   &    ...&    40.6&     55.5&      67.4&      60.5&     67.9&     50.9&      54.8&      42.4&     54.5&     52.4&    51.8&     47.9\\
4883.69&    Y~{\sc ii}&    1.08&   0.070&   &   41.4&    41.0&     37.2&      56.8&      82.3&     73.4&     58.9&      55.0&      56.3&     53.1&     58.9&    64.9&     47.1\\
4900.11&    Y~{\sc ii}&    1.03&  -0.090&   &   80.2&     ...&      ...&      65.6&       ...&    160.8&      ...&       ...&       ...&     64.1&     77.2&    81.4&     53.8\\
5087.43&    Y~{\sc ii}&    1.08&  -0.170&   &    ...&    32.9&     29.2&      66.4&      38.9&     64.0&     47.0&      30.2&      42.0&     36.3&     39.8&    47.6&     34.9\\
5200.42&    Y~{\sc ii}&    0.99&  -0.570&   &   34.5&    29.9&      ...&      49.4&      40.9&     53.5&      ...&       ...&      36.1&     23.4&      ...&     ...&     20.0\\
4810.54&    Zn~{\sc i}&    4.08&  -0.170&  *&  22.0 &   40.9 &    31.9 &     32.9 &     43.0 &    50.0 &    36.9 &     34.8 &     36.6 &     30.2&     33.2&    30.9&     26.4\\
\end{longtable}
\end{center}
\end{landscape}
\clearpage

\end{singlespace}

\begin{singlespace}
\begin{table*}[!hbt]
\begin{center}
\caption{Fornax elemental ratios \label{tab:ratios_fnx}}
\begin{tabular}{r|rrr|rrr|rrr|}
\hline 
\hline 
\multicolumn{10}{c}{}\\
\cline{2-10}
&    Cl1-D56& $\sigma$& N$_{\mathrm{lines}}$&  Cl1-D68& $\sigma$& N$_{\mathrm{lines}}$&  Cl1-D164& $\sigma$& N$_{\mathrm{lines}}$\\
\hline
$\textrm{[Ba~{\sc ii}/Fe~{\sc i}]}$&        -0.13&   0.06&    4&        0.06&   0.09&    4&      0.07&   0.07&    4   \\
$\textrm{[Ca~{\sc i}/Fe~{\sc i}]}$&        0.27&   0.09&    5&        0.18&   0.07&    5&      0.09&   0.05&    4   \\
$\textrm{[Cr~{\sc i}/Fe~{\sc i}]}$&        -0.36&   0.26&    2&        -0.35&   0.12&    2&      -0.43&   0.16&    2   \\
$\textrm{[Eu~{\sc ii}/Fe~{\sc i}]}$&          ...&    ...&    0&     $<$1.04&   0.00&    1&   $<$0.89&   0.00&    1   \\
$\textrm{[Fe~{\sc i}/H]}$&        -2.40&   0.03&   40&        -2.55&   0.03&   39&      -2.59&   0.03&   45   \\
$\textrm{[Fe~{\sc ii}/H]}$&        -2.43&   0.06&    4&        -2.62&   0.11&    5&      -2.56&   0.06&    7   \\
$\textrm{[La~{\sc ii}/Fe~{\sc i}]}$&          ...&    ...&    0&          ...&    ...&    0&        ...&    ...&    0   \\
$\textrm{[Mg~{\sc i}/Fe~{\sc i}]}$&        0.52&   0.12&    3&        0.38&   0.07&    3&      0.08&   0.07&    3   \\
$\textrm{[Mn~{\sc i}/Fe~{\sc i}]}$&          ...&    ...&    0&          ...&    ...&    0&        ...&    ...&    0   \\
$\textrm{[Na~{\sc i}/Fe~{\sc i}]}$&        -0.10&   0.12&    2&        -0.15&   0.08&    2&      0.42&   0.12&    2   \\
$\textrm{[Nd~{\sc ii}/Fe~{\sc i}]}$&        0.60&   0.20&    1&        0.40&   0.07&    2&      0.49&   0.10&    1   \\
$\textrm{[Ni~{\sc i}/Fe~{\sc i}]}$&          ...&    ...&    0&        0.27&   0.23&    1&      -0.20&   0.21&    1   \\
$\textrm{[O~{\sc i}/Fe~{\sc i}]}$&     $<$0.68&   0.00&    1&        0.28&   0.10&    1&      0.37&   0.10&    1   \\
$\textrm{[Ti~{\sc i}/Fe~{\sc i}]}$&        0.07&   0.12&    5&        -0.22&   0.09&    5&      -0.09&   0.14&    5   \\
$\textrm{[Ti~{\sc ii}/Fe~{\sc i}]}$&        0.06&   0.10&    5&        -0.02&   0.06&    6&      0.18&   0.08&    5   \\
$\textrm{[Y~{\sc ii}/Fe~{\sc i}]}$&        -0.21&   0.28&    2&        -0.29&   0.14&    3&      -0.33&   0.26&    2   \\
$\textrm{[Zn~{\sc i}/Fe~{\sc i}]}$&     $<$-0.10&   0.00&    1&     $<$0.45&   0.00&    1&   $<$0.29&   0.00&    1   \\
\hline 
\multicolumn{10}{c}{}\\
\cline{2-10}
&    Cl2-B71& $\sigma$& N$_{\mathrm{lines}}$&  Cl2-B77& $\sigma$& N$_{\mathrm{lines}}$&  Cl2-B226& $\sigma$& N$_{\mathrm{lines}}$\\
\hline
$\textrm{[Ba~{\sc ii}/Fe~{\sc i}]}$&        -0.19&   0.10&    4&        -0.12&   0.10&    3&      -0.38&   0.10&    3   \\
$\textrm{[Ca~{\sc i}/Fe~{\sc i}]}$&        0.27&   0.07&    7&        0.16&   0.05&    6&      0.21&   0.04&    9   \\
$\textrm{[Cr~{\sc i}/Fe~{\sc i}]}$&        -0.38&   0.19&    2&        -0.10&   0.16&    2&      -0.03&   0.22&    2   \\
$\textrm{[Eu~{\sc ii}/Fe~{\sc i}]}$&     $<$0.63&   0.00&    1&     $<$0.88&   0.00&    1&   $<$0.60&   0.00&    1   \\
$\textrm{[Fe~{\sc i}/H]}$&        -2.14&   0.03&   50&        -2.09&   0.04&   47&      -2.01&   0.02&   47   \\
$\textrm{[Fe~{\sc ii}/H]}$&        -2.06&   0.06&    6&        -2.03&   0.07&    6&      -2.01&   0.04&    8   \\
$\textrm{[La~{\sc ii}/Fe~{\sc i}]}$&          ...&    ...&    0&          ...&    ...&    0&        ...&    ...&    0   \\
$\textrm{[Mg~{\sc i}/Fe~{\sc i}]}$&        0.53&   0.08&    3&        0.43&   0.08&    3&      0.28&   0.07&    2   \\
$\textrm{[Mn~{\sc i}/Fe~{\sc i}]}$&          ...&    ...&    0&          ...&    ...&    0&      -0.28&   0.07&    2   \\
$\textrm{[Na~{\sc i}/Fe~{\sc i}]}$&        -0.08&   0.09&    2&        -0.25&   0.12&    2&        ...&    ...&    0   \\
$\textrm{[Nd~{\sc ii}/Fe~{\sc i}]}$&          ...&    ...&    0&          ...&    ...&    0&      -0.13&   0.12&    2   \\
$\textrm{[Ni~{\sc i}/Fe~{\sc i}]}$&        0.09&   0.29&    2&        -0.01&   0.25&    1&      0.09&   0.26&    2   \\
$\textrm{[O~{\sc i}/Fe~{\sc i}]}$&        0.37&   0.15&    1&        0.32&   0.20&    1&      0.49&   0.10&    1   \\
$\textrm{[Ti~{\sc i}/Fe~{\sc i}]}$&        -0.03&   0.13&    5&        -0.13&   0.10&    4&      -0.08&   0.06&    5   \\
$\textrm{[Ti~{\sc ii}/Fe~{\sc i}]}$&        0.05&   0.07&    5&        0.18&   0.09&    4&      0.16&   0.06&    5   \\
$\textrm{[Y~{\sc ii}/Fe~{\sc i}]}$&        -0.10&   0.15&    4&        -0.44&   0.23&    2&      -0.25&   0.18&    3   \\
$\textrm{[Zn~{\sc i}/Fe~{\sc i}]}$&        -0.11&   0.20&    1&        0.09&   0.20&    1&      0.11&   0.20&    1   \\
\hline 
\multicolumn{10}{c}{}\\
\cline{2-10}
&    Cl3-B59& $\sigma$& N$_{\mathrm{lines}}$&  Cl3-B61& $\sigma$& N$_{\mathrm{lines}}$&  Cl3-B82& $\sigma$& N$_{\mathrm{lines}}$\\
\hline
$\textrm{[Ba~{\sc ii}/Fe~{\sc i}]}$&        0.27&   0.09&    3&        0.09&   0.10&    3&      0.27&   0.09&    2   \\
$\textrm{[Ca~{\sc i}/Fe~{\sc i}]}$&        0.27&   0.17&    4&        0.21&   0.03&    6&      0.23&   0.05&    5   \\
$\textrm{[Cr~{\sc i}/Fe~{\sc i}]}$&        -0.30&   0.20&    2&        -0.28&   0.20&    2&      -0.50&   0.15&    2   \\
$\textrm{[Eu~{\sc ii}/Fe~{\sc i}]}$&        0.89&   0.10&    1&        0.78&   0.15&    1&      0.97&   0.10&    1   \\
$\textrm{[Fe~{\sc i}/H]}$&        -2.35&   0.02&   44&        -2.42&   0.03&   50&      -2.38&   0.03&   48   \\
$\textrm{[Fe~{\sc ii}/H]}$&        -2.30&   0.09&    5&        -2.31&   0.09&    4&      -2.36&   0.06&    8   \\
$\textrm{[La~{\sc ii}/Fe~{\sc i}]}$&     $<$0.52&   0.04&    5&        0.95&   0.10&    1&      0.62&   0.06&    3   \\
$\textrm{[Mg~{\sc i}/Fe~{\sc i}]}$&        0.19&   0.09&    3&        0.37&   0.07&    3&      -0.35&   0.08&    3   \\
$\textrm{[Mn~{\sc i}/Fe~{\sc i}]}$&          ...&    ...&    0&        0.03&   0.07&    2&      -0.01&   0.07&    2   \\
$\textrm{[Na~{\sc i}/Fe~{\sc i}]}$&        0.05&   0.14&    2&        -0.25&   0.12&    2&      0.48&   0.13&    2   \\
$\textrm{[Nd~{\sc ii}/Fe~{\sc i}]}$&        0.65&   0.07&    2&        0.44&   0.09&    2&      0.73&   0.07&    2   \\
$\textrm{[Ni~{\sc i}/Fe~{\sc i}]}$&        0.22&   0.27&    1&        -0.04&   0.24&    1&      -0.02&   0.23&    1   \\
$\textrm{[O~{\sc i}/Fe~{\sc i}]}$&        0.43&   0.10&    1&        0.65&   0.10&    1&      0.16&   0.10&    1   \\
$\textrm{[Ti~{\sc i}/Fe~{\sc i}]}$&        0.02&   0.08&    6&        0.04&   0.09&    4&      -0.17&   0.07&    5   \\
$\textrm{[Ti~{\sc ii}/Fe~{\sc i}]}$&        0.18&   0.05&    6&        0.30&   0.08&    6&      0.05&   0.03&    5   \\
$\textrm{[Y~{\sc ii}/Fe~{\sc i}]}$&        -0.23&   0.30&    2&        -0.19&   0.19&    2&      -0.24&   0.18&    3   \\
$\textrm{[Zn~{\sc i}/Fe~{\sc i}]}$&     $<$0.15&   0.00&    1&     $<$0.22&   0.00&    1&      0.18&   0.10&    1   \\
\hline 
\multicolumn{10}{c}{}\\
\hline 
\hline 
\end{tabular}
\end{center}
\end{table*}

\end{singlespace}

\begin{singlespace}
\begin{table*}[!hbt]
\begin{center}
\caption{M15 elemental ratios \label{tab:ratios_m15}}
\begin{tabular}{r|rrr|rrr|}
\hline 
\hline 
\multicolumn{7}{c}{}\\
\cline{2-7}
&    M15S1& $\sigma$& N$_{\mathrm{lines}}$&  M15S3& $\sigma$& N$_{\mathrm{lines}}$\\
\hline
$\textrm{[Ba~{\sc ii}/Fe~{\sc i}]}$&        -0.15&   0.03&    4&        0.29&   0.06&    4   \\
$\textrm{[Ca~{\sc i}/Fe~{\sc i}]}$&        0.32&   0.03&    7&        0.37&   0.03&    7   \\
$\textrm{[Cr~{\sc i}/Fe~{\sc i}]}$&        -0.43&   0.05&    2&        -0.33&   0.05&    2   \\
$\textrm{[Eu~{\sc ii}/Fe~{\sc i}]}$&        0.30&   0.20&    1&        0.65&   0.10&    1   \\
$\textrm{[Fe~{\sc i}/H]}$&        -2.36&   0.02&   55&        -2.41&   0.02&   52   \\
$\textrm{[Fe~{\sc ii}/H]}$&        -2.37&   0.03&   12&        -2.35&   0.04&   12   \\
$\textrm{[La~{\sc ii}/Fe~{\sc i}]}$&        0.09&   0.10&    1&        0.36&   0.07&    3   \\
$\textrm{[Mg~{\sc i}/Fe~{\sc i}]}$&        0.33&   0.07&    3&        0.28&   0.07&    3   \\
$\textrm{[Mn~{\sc i}/Fe~{\sc i}]}$&        -0.59&   0.05&    5&        -0.38&   0.07&    6   \\
$\textrm{[Na~{\sc i}/Fe~{\sc i}]}$&        0.03&   0.14&    2&        0.68&   0.11&    2   \\
$\textrm{[Nd~{\sc ii}/Fe~{\sc i}]}$&        -0.14&   0.12&    1&        0.30&   0.04&    2   \\
$\textrm{[Ni~{\sc i}/Fe~{\sc i}]}$&        0.07&   0.06&    3&        0.11&   0.06&    3   \\
$\textrm{[O~{\sc i}/Fe~{\sc i}]}$&        0.04&   0.10&    1&          ...&    ...&    0   \\
$\textrm{[Ti~{\sc i}/Fe~{\sc i}]}$&        -0.08&   0.09&    5&        -0.06&   0.07&    5   \\
$\textrm{[Ti~{\sc ii}/Fe~{\sc i}]}$&        0.13&   0.05&    5&        0.16&   0.07&    5   \\
$\textrm{[Y~{\sc ii}/Fe~{\sc i}]}$&        -0.42&   0.02&    4&        -0.20&   0.04&    3   \\
$\textrm{[Zn~{\sc i}/Fe~{\sc i}]}$&        0.03&   0.05&    1&        0.15&   0.09&    1   \\
\hline 
\multicolumn{7}{c}{}\\
\cline{2-7}
&    M15S6& $\sigma$& N$_{\mathrm{lines}}$&  M15S7& $\sigma$& N$_{\mathrm{lines}}$\\
\hline
$\textrm{[Ba~{\sc ii}/Fe~{\sc i}]}$&        0.39&   0.04&    4&        -0.20&   0.06&    4   \\
$\textrm{[Ca~{\sc i}/Fe~{\sc i}]}$&        0.33&   0.03&    6&        0.41&   0.03&    7   \\
$\textrm{[Cr~{\sc i}/Fe~{\sc i}]}$&        -0.39&   0.06&    2&        -0.39&   0.04&    2   \\
$\textrm{[Eu~{\sc ii}/Fe~{\sc i}]}$&        0.81&   0.10&    1&        0.41&   0.10&    1   \\
$\textrm{[Fe~{\sc i}/H]}$&        -2.32&   0.03&   56&        -2.47&   0.04&   49   \\
$\textrm{[Fe~{\sc ii}/H]}$&        -2.35&   0.06&   13&        -2.50&   0.04&   11   \\
$\textrm{[La~{\sc ii}/Fe~{\sc i}]}$&        0.37&   0.12&    2&        0.09&   0.18&    2   \\
$\textrm{[Mg~{\sc i}/Fe~{\sc i}]}$&        -0.15&   0.07&    3&        0.43&   0.07&    3   \\
$\textrm{[Mn~{\sc i}/Fe~{\sc i}]}$&        -0.41&   0.06&    5&        -0.45&   0.20&    1   \\
$\textrm{[Na~{\sc i}/Fe~{\sc i}]}$&        0.59&   0.14&    2&        0.08&   0.12&    2   \\
$\textrm{[Nd~{\sc ii}/Fe~{\sc i}]}$&        0.28&   0.03&    2&        -0.23&   0.16&    1   \\
$\textrm{[Ni~{\sc i}/Fe~{\sc i}]}$&        0.06&   0.06&    3&        0.00&   0.06&    3   \\
$\textrm{[O~{\sc i}/Fe~{\sc i}]}$&        -0.05&   0.10&    1&        0.15&   0.10&    1   \\
$\textrm{[Ti~{\sc i}/Fe~{\sc i}]}$&        -0.09&   0.07&    6&        -0.04&   0.11&    4   \\
$\textrm{[Ti~{\sc ii}/Fe~{\sc i}]}$&        0.14&   0.05&    5&        0.07&   0.07&    5   \\
$\textrm{[Y~{\sc ii}/Fe~{\sc i}]}$&        -0.15&   0.04&    3&        -0.37&   0.04&    4   \\
$\textrm{[Zn~{\sc i}/Fe~{\sc i}]}$&        0.02&   0.10&    1&        0.04&   0.08&    1   \\
\hline 
\multicolumn{7}{c}{}\\
\hline 
\hline 
\end{tabular}
\end{center}
\end{table*}

\end{singlespace}

\end{document}